\documentclass[aps,pre,citeautoscript,twocolumn,superscriptaddress,nofootinbib,nobibnotes,10pt]{revtex4-2} 
\usepackage[utf8]{inputenc}
\usepackage{amsmath,amssymb}
\usepackage{mathrsfs} 
\usepackage{mathtools}
\usepackage{siunitx} 
\usepackage{graphicx,color}
\usepackage{varwidth,xcolor}
\usepackage{array} 
\usepackage{placeins}
\usepackage{ifthen}
\usepackage{hyperref}\usepackage[capitalise]{cleveref}
\usepackage{natbib} 
\usepackage[normalem]{ulem} 

\DeclareGraphicsExtensions{.pdf,.PDF,.png,.PNG}

\definecolor{myblue}{RGB}{20,80,150} 	
\definecolor{myred}{RGB}{160,30,30} 	
\definecolor{mygreen}{RGB}{50,120,50} 	

\hypersetup{
	colorlinks=true,
	linkcolor=myblue, 	
	citecolor=myred, 	
	urlcolor=mygreen 	
}

\newcommand{\dif}{\mathrm{d}}			
\newcommand{\id}{\operatorname{id}}		

\newcommand{\ii}{\mathrm{i}}			
\newcommand{\ee}{\mathrm{e}}			
\newcommand{\tdif}[2]{\frac{\dif#1}{\dif#2}} 
\newcommand{\pdif}[2]{\frac{\partial#1}{\partial#2}} 
\newcommand{\N}{\mathbb{N}}				
\newcommand{\abs}[1]{\lvert#1\rvert}   	
\newcommand{\average}[1]{\langle#1\rangle} 
\newcommand{\vectwo}[2]{\begin{pmatrix} #1 \\ #2 \end{pmatrix}} 
\newcommand{\mat}[1]{\mathbf{#1}}		
\newcommand{\arctanh}{\operatorname{arctanh}} 
\DeclareMathOperator{\sign}{sign} 		
\newcommand{\E}{\mathbb{E}}				
\newcommand{\Var}{\mathbb{V}}			
\newcommand{\floor}[1]{\lfloor #1 \rfloor} 

\newcommand{\sub}[2]{#1_\mathrm{#2}} 	

\newlength{\myl}%
\newcommand{\SUM}[2]{{\setlength{\myl}{\widthof{$\displaystyle\sum_{#1}^{#2}$}*\real{0.5}-\widthof{$\displaystyle\sum$}*\real{0.5}}\sum_{#1}^{#2}\;\hspace{-\the\myl}}} 
\newcommand{\INT}[3]{\settowidth{\myl}{$\displaystyle\int_{#1}^{#2}$}{\int_{#1}^{#2}\;\;\:\hspace{-\the\myl}\dif #3}\,}
\newcommand{\TINT}[3]{\settowidth{\myl}{$\int_{#1}^{#2}$}{\int_{#1}^{#2}\!\ifthenelse{\equal{#1#2}{}}{}{\;\;\;\;\hspace{-\the\myl}}\dif #3}\,}%
\newcommand{\EINT}[3]{\settowidth{\myl}{$\int_{#1}^{#2}$}{\int_{#1}^{#2}\;\;\;\,\hspace{-\the\myl}\dif #3}\,}

\begin{document}
	
	\title{Excitability and memory in a time-delayed optoelectronic neuron}
	
	\author{Jonas Mayer Martins}
	\email[Corresponding author: ]{jonas.mayermartins@uni-muenster.de}
	\affiliation{Institute for Theoretical Physics, University of M\"unster, Wilhelm-Klemm-Str.\ 9 \\ and Center for Nonlinear Science (CeNoS), University of M\"unster, Corrensstrasse 2, 48149 M\"unster, Germany}
	
	\author{Svetlana V. Gurevich}
	\affiliation{Institute for Theoretical Physics, University of M\"unster, Wilhelm-Klemm-Str.\ 9 \\ and Center for Nonlinear Science (CeNoS), University of M\"unster, Corrensstrasse 2, 48149 M\"unster, Germany}
	
	\author{Julien Javaloyes}
	\affiliation{Departament de F\'isica and IAC-3, Universitat de les Illes Balears, C/ Valldemossa km 7.5, 07122 Palma de Mallorca, Spain}

	\begin{abstract}
		We study the dynamics of an optoelectronic circuit composed of an excitable nanoscale resonant-tunneling diode~(RTD) driving a nanolaser diode~(LD) coupled via time-delayed feedback. Using a combination of numerical path-continuation methods and time simulations, we demonstrate that this RTD-LD system can serve as an artificial neuron, generating pulses in the form of temporal localized states (TLSs) that can be employed as memory for neuromorphic computing. In particular, our findings reveal that the prototypical delayed FitzHugh--Nagumo model previously employed to model the RTD-LD resembles our more realistic model only in the limit of a slow RTD. We show that the RTD time scale plays a critical role in memory capacity as it governs a shift in pulse interaction from repulsive to attractive, leading to a transition from stable to unstable multi-pulse TLSs. Our theoretical analysis uncovers features and challenges previously unknown for the RTD-LD system, including the multistability of TLSs and attractive interaction forces, stemming from the previously neglected intrinsic dynamics of the laser. These effects are crucial to consider since they define the memory properties of the RTD-LD.
	\end{abstract}
	\maketitle
	
	\section{\label{sec:introduction}Introduction}
	
	The human brain is arguably the most exciting matter in the universe. Consuming as little power as a light bulb, the brain is extremely power-efficient and still outperforms artificial computers in many ways~\cite{CowleyEtAl2022}.
	The vast majority of modern-day computers implement the von Neumann architecture~\cite{vN1993}.
	Considering the scientific, technological, and socio-cultural progress that computers have bestowed upon us~\cite{HobartS2000}, the classical computing architecture has served us well.
	Yet with the ever-increasing demand for higher computing power and the advent of artificial intelligence, major issues have become apparent. First, classical computers encode information digitally, which entails high energy consumption~\cite{FreitagBLWKBF2021}, primarily due to heat dissipation. Second, the CPU processes information sequentially, limiting bandwidth. Furthermore, the physical distance between computational units slows down computation even further. Last, the size of transistors, which primarily drives Moore's law, is limited by quantum effects~\cite{VREtAl2003}.
	
	The answer to these challenges might be to mimic the brain. So-called \emph{neuromorphic computing} emulates the structure of the brain by connecting artificial neurons in a network, thus merging memory and processing units~\cite{MarkovicMQG2020}. Neuromorphic computing is particularly suited for implementing integrated machine-learning algorithms~\cite{ShastriTdLPBWP2021}. An essential property of neurons allowing them to process and transmit information is \emph{excitability}~\cite{GerstnerK2002}.
	From a dynamical systems perspective~\cite{Izhikevich1999,Izhikevich2007}, a system is excitable if a sufficiently strong perturbation of the resting state elicits a large-amplitude excursion in phase space that is largely independent of the details of the perturbation and subsequently returns to the resting state. For neurons, this large-amplitude response of the membrane potential is called an \emph{action potential}, \emph{pulse}, or \emph{spike}. A prototypical model of excitability is the FitzHugh--Nagumo (FHN) neuron~\cite{BukhSEMSS2023,Izhikevich2007,LindnerGONSG2004,ZakharovaSAS2017,MarinoG2018,MarinoG2019}. 
	
	Excitability is a ubiquitous concept, not only in biology~\cite{Murray2002,Murray2003,QuHGW2014} but also in chemistry, e.g., the Belousov--Zhabotinsky reaction~\cite{KuhnertAK1989,Murray2002}, and in physics, e.g., lasers~\cite{SelmiBBSKB2014,ShastriNTRWP2016}, particles trapped in an optical torque wrench~\cite{PedaciHvOBD2010}, and resonant-tunneling diodes (RTDs)~\cite{RomeiraFJ2017}. 
	The combination of excitability and delay is known to give rise to \emph{temporal localized states} (TLSs) in many systems, such as in a semiconductor laser with coherent optical injection~\cite{GarbinJTB2015,MunsbergJG2020}, the FHN neuron with delayed
	feedback~\cite{RomeiraAFBJ2016,YanchukRSW2019}, and the Morris--Lecar model of biological neurons~\cite{MorrisL1981,TsukumaK2020,StoehrW2023}. Since TLSs are self-healing in the sense that these solutions are robust to perturbations, they can encode information and act as stable memory.
	
	A multitude of electronic excitable systems have been studied as potential candidates for artificial neurons and tested experimentally on neuromorphic chips, e.g., Neurogrid~\cite{BenjaminEtAl2014}, TrueNorth~\cite{MerollaEtAl2014}, as well as SpiNNaker~\cite{FurberGTP2014} and FACETS~\cite{SchemmelBGHMM2010} as part of the Human Brain Project~\cite{Markram2012,LorentsEtAl2023}.
	Electronic artificial neurons, however, are relatively slow ($\mathrm{kHz}$) and suffer from heat loss due to dissipation in electric interconnects, which makes them energy-intensive ($\mathrm{pJ/spike}$)~\cite{Miller2017}.
	Conversely, optical computing promises high computing speeds at extremely low energy costs~\cite{CaulfieldD2010,FdLSTNP2017,ShastriTdLPBWP2021}. Optical or optoelectronic artificial neurons have been implemented, for example, as a semiconductor ring laser~\cite{CoomansGBDVdS2011}, graphene excitable laser~\cite{ShastriNTRWP2016}, time-delayed optoelectronic resonator~\cite{RomeiraJIFBP2013,RomeiraAFBJ2016}, and vertical-cavity surface-emitting laser (VCSEL)~\cite{MarinoG2019,HejdaRBAH2021}.
	Miniaturization of the devices is another avenue to further increase efficiency. Although smaller sizes come with challenges associated with the diffraction limit and other quantum effects, such as the Purcell effect~\cite{RomeiraF2018}, nanoscale devices, e.g., semiconductor lasers, promise high speed and require little power~\cite{HillG2014,MaO2018}.
	
	In this article, we study a system consisting of a \emph{resonant-tunneling diode} (RTD) driving a \emph{laser diode} (LD) subjected to time-delayed feedback. Both the RTD and the LD can be of nanoscopic scale in our model. We derive the stochastic delay-differential equations (DDEs) that describe the combined RTD-LD system from Ref.~\cite{OPHALHFRJ2022}, where this system has been shown numerically to be excitable and propagate pulses from one neuron to another.
	While there has been a preliminary experimental implementation of the RTD-LD~\cite{RomeiraJIFBP2013}, which demonstrated the feasibility of the device, this implementation had its limitations because the device was operated at a low frequency and was used only to analyze single-pulse dynamics without time-delayed feedback.
	Yet an experimental setup operating at a higher frequency -- on the order of several GHz~\cite{RomeiraFJ2017} -- is possible. Our present analysis paves the way for a more sophisticated experimental implementation.
	Furthermore, from a theoretical perspective, the simple FHN model with time delay, as discussed in Ref.~\cite{RomeiraAFBJ2016}, cannot reproduce the complexity due to the competition between the time scales of the laser and the RTD. With the more realistic RTD-LD model presented here, we achieve a complete understanding of these features.
	
	With this motivation, we perform a comprehensive theoretical analysis using a combination of time simulations and path-continuation methods to determine how the RTD-LD can function as an artificial neuron. Most importantly, we shall demonstrate that our model exhibits TLSs as solutions and discuss under which conditions the paradigmatic FHN model, which was employed in an earlier theoretical study of the RTD-LD~\cite{RomeiraAFBJ2016} and neglected the laser dynamics through an adiabatic approximation and Pyragas-type feedback~\cite{Pyragas1992,SchoellS2007}, is justified. We show that the DDE model employed here qualitatively reproduces the delayed FHN model in a limiting case, yet exhibits new features and challenges -- multistability of TLSs and instability of multi-pulse states due to attractive interaction forces -- that arise due to the nanoscale laser.
	
	This paper is structured as follows: First, we introduce the DDEs that model the RTD-LD along with the numerical methods in \cref{sec:methods}. Subsequently, we analyze the system by combining time simulations and path continuation methods in \cref{sec:results}. This analysis consists of four main parts: a brief overview of the RTD without feedback (as studied in Ref.~\cite{OPPFRJ2021}) in \cref{sec:results_noFeedback}, then the RTD-LD subject to feedback with a slow RTD in \cref{sec:results_slowRTD} and a fast RTD in \cref{sec:results_fastRTD} as well as a discussion of the characteristic time scale connecting these two regimes in \cref{sec:results_time-scale}. Finally, we discuss our results, in particular the impact of the RTD time scale on the memory properties of the system and hence its viability as an artificial neuron, and give an outlook in \cref{sec:conclusions}.

	\begin{figure*}[tbhp]
		\centering
		\includegraphics[width=0.99\linewidth]{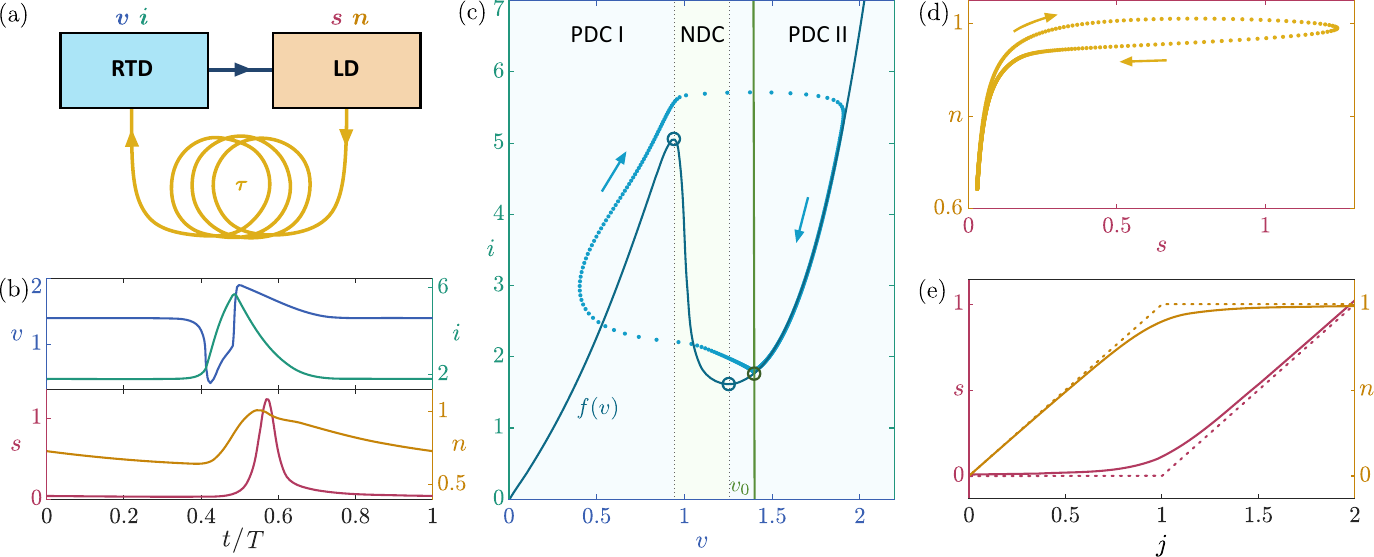}
		\caption{(a) Schematic of the RTD-LD with feedback of time delay $\tau$. The dynamic variables are the voltage~$v$ and the current~$i$ for the RTD, as well as the photon number~$s$ and the carrier number~$n$ for the LD. (b) Time trace of an excited pulse at bias voltage $v_0 = 1.4$. (c) Phase space of the RTD: Nonlinear current-voltage characteristic~$f(v)$ of the RTD-LD (dark blue line), local extrema of $f(v)$ (dark blue circles), load line $v_0 = v - ri$ (green line), and the resulting operating point (green circle), along with the excited pulse from (b) (light blue), PDC regions (blue regions), and NDC region (green region). (d) Phase space of the LD with the excited pulse from (b). (e) Equilibrium states of the LD depending on the bias current~$j$ for $\eta = 0$. The exact solution (solid line) approximates the transcritical bifurcation (dotted line) for $\sub{\gamma}{m} / \sub{\gamma}{t} \to 0$.}
		\label{fig:1}
	\end{figure*}

	\section{\label{sec:methods}Model system and methods}
	
	\subsection{\label{sec:methods_intro_to_RTD-LD}The RTD-LD model}
	
	The RTD-LD is an optoelectronic circuit capable of self-oscillation, generating pulses from perturbations, and sustaining these pulses through time-delayed feedback. In~\cref{fig:1}~(a), we illustrate the basic operating principle: Here, the RTD oscillates in the voltage~$v$ and the current~$i$, which drives the laser carrier number~$n$. In turn, the carrier number interacts with the photon number $s$ through spontaneous and stimulated emission. The resulting light pulse travels back to the RTD via, e.g., an optical fiber, which induces a delay time~$\tau$. Such a pulse, as exemplified in panel~(b), is a clockwise orbit in the phase space of the RTD in panel~(c) as well as that of the laser in panel~(d). Once the delayed light pulse reaches the RTD, it results in a photocurrent that affects the voltage~$v$. This time-delayed feedback renders the RTD-LD an \emph{autaptic} neuron~\cite{BacciHP2003,GuoEtAl2016}, akin to a chain of identical neurons, which allows us to study information propagation and memory with much lower computational cost.
	
	The model of the RTD-LD comprises a set of four DDEs in time $t$ for the voltage $v(t)$, the current $i(t)$, the photon number $s(t)$, and the carrier number $n(t)$ 
	\begin{align}
		\sub{t}{v}\hspace{1.5pt} \dot{v} &= i - f(v) - \kappa s(t - \tau)
		\,,\label{eq:system_v}\\
		\sub{t}{i}\hspace{5pt}   \dot{i} &= v_{0} - v - ri
		\,,\label{eq:system_i}\\
		\sub{t}{s}\hspace{2.5pt} \dot{s} &= (n - 1)s + \frac{\sub{\gamma}{m}}{\sub{\gamma}{t}} (n_{0} + n)
		\,,\label{eq:system_s}\\
		\sub{t}{n}               \dot{n} &= j + \eta i - n(1 + s)
		\,,\label{eq:system_n}
	\end{align}
	see \cref{app:sec:model-derivation} for the derivation and variable scaling. The parameters, listed in \cref{app:sec:typical_parameters} with typical values, are the current-voltage characteristic~$f(v)$, the feedback strength~$\kappa$ from the LD to the RTD with delay~$\tau$, the bias voltage~$v_0$, the resistance~$r$, the spontaneous emission into the lasing mode~$\sub{\gamma}{m}$ and the total decay rate~$\sub{\gamma}{t}$, the transparency carrier number~$n_0$, the injection efficiency~$\eta$ of the RTD into the LD, and finally the bias current~$j$. Time is normalized to the characteristic time scale $\sub{t}{c}$, and each of the variables $v$, $i$, $s$, and $n$ has its own characteristic time scale $\sub{t}{v}$, $\sub{t}{i}$, $\sub{t}{s}$, and $\sub{t}{n}$, respectively. Typically, $\sub{t}{s}\ll \sub{t}{n}$ and $\sub{t}{v}\ll \sub{t}{i}$ so that both the RTD and the LD are slow-fast systems. This is evident from the time trace of a typical periodic solution of period~$T$ shown in~\cref{fig:1}~(b).
	
	Note that the RTD is a slow-fast system based on resonant-tunneling through a double-quantum well~\cite{ChangET1974}: While the voltage $v$ changes quickly on the time scale~$\sub{t}{v}$, the current $i$ follows slowly on the time scale $\sub{t}{i}$, see \cref{fig:1}~(b) and (c).
	From~\cref{eq:system_v,eq:system_i}, we can easily deduce that the steady state of the RTD lies at the intersection of the nullclines $i = f(v)$ and $i = (v_0 - v)/r$. Assuming a very small resistance $r$, the slope of the latter nullcline is nearly vertical, which guarantees that for every bias voltage $v_0$, there is only one intersection point with $f(v)$ at $v \approx v_0$, as shown in~\cref{fig:1}~(c). The stability of this fixed point, however, varies with the bias voltage $v_0$, which is another important parameter determining the RTD-LD dynamics.
	
	Without feedback, i.e., for $\kappa = 0$, the RTD Eqs.~\eqref{eq:system_v} and \eqref{eq:system_i} are equivalent to the classic FitzHugh--Nagumo (FHN) model~\cite{FitzHugh1961,NagumoAY1962}, except that the current-voltage characteristic $f(v)$ is not the cubic polynomial $-v + v^3 / 3$ but the function
	\begin{equation}
		\begin{split}
			f(v) &= \sign(a) \log
			\!\Bigg(\frac{
				1 + \exp\!\big(
				\frac{q_\mathrm{e}}{k_\mathrm{B} T}
				(b - c + n_1 v)
				\big)
			}{
				1 + \exp\!\big(
				\frac{q_\mathrm{e}}{k_\mathrm{B} T}
				(b - c - n_1 v)
				\big)
			}\Bigg)
			\\&\quad \times
			\bigg[
			\frac{\pi}{2}
			+ \arctan\!\bigg(\frac{c - n_1 v}{d}\bigg)
			\bigg]
			\\&\quad
			+ \frac{h}{\abs{a}}
			\bigg[
			\exp\!\bigg(\frac{q_\mathrm{e}}{k_\mathrm{B} T} n_2 v\bigg)	- 1
			\bigg],
		\end{split}
		\label{eq:current-voltage}
	\end{equation}
	where $a$, $b$, $c$, $d$, $h$, $n_1$, and $n_2$ are fit parameters, $\sub{q}{e}$ is the elementary charge, $\sub{k}{B}$ is the Boltzmann constant, and $T$ is the temperature.
	This expression is derived in Ref.~\cite{SchulmanDLSC1996} by mixing first-principle calculations with a fit of experimental data of RTDs. The slope of $f(v)$ is the differential conductance and, for typical parameters (see \cref{app:sec:typical_parameters}), the characteristic has a region of \emph{negative} differential conductance (NDC) in between two regions of \emph{positive} differential conductance (PDC~I and PDC~II), cf.~\cref{fig:1}~(c). The NDC region is the key property of RTDs and precisely in this region, the steady state loses stability, leading to self-oscillation, as we shall discuss in more detail in~\cref{sec:results_noFeedback}.
	
	The LD, being a class-B laser, is a slow-fast system too, see~\cref{fig:1}~(b) and (d): The photon number~$s$ is fast on the time scale $\sub{t}{s} = \sub{\tau}{s} / \sub{t}{c}$ with photon lifetime~$\sub{\tau}{s}$. Once excited by the current~$i$ with an efficiency~$\eta$, the slow carrier number~$n$ returns to equilibrium with an exponential decay on the time scale $\sub{t}{n} = \sub{\tau}{n} / \sub{t}{c}$, where $\sub{\tau}{n} = 1 / \sub{\gamma}{t}$ is the carrier lifetime. Conversely, $\sub{\gamma}{t} = \sub{\gamma}{l}+\sub{\gamma}{m}+\sub{\gamma}{nr}$ is the total decay rate, consisting of the spontaneous emission in the leaky modes $\sub{\gamma}{l}$ and the lasing mode~$\sub{\gamma}{m}$, as well as the non-radiative recombination $\sub{\gamma}{nr}$. The ratio $\sub{\gamma}{m} / \sub{\gamma}{t} = \beta \mathrm{QE}$ is the product of the spontaneous coupling rate $\beta$~\cite{RomeiraF2018} and the quantum efficiency $\mathrm{QE}$ (cf.~\cref{app:eq:spont_em_coupling_factor_definition,app:eq:QE_definition}). We choose the bias parameter~$j \approx -0.43$ in the off-state near an approximate transcritical bifurcation at $j = 1$ (which is exactly a transcritical bifurcation for $\sub{\gamma}{m} / \sub{\gamma}{t} \to 0$) so that the laser turns on intermittently when driven by the RTD, see~\cref{fig:1}~(e).

	\subsection{\label{sec:methods_slow-RTD_approximation}Slow-RTD approximation}
	
	Our analysis of the RTD-LD system~\mbox{\eqref{eq:system_v}-\eqref{eq:system_n}} focuses on the impact of the laser nonlinearity and the RTD time scale on the behavior of TLSs. When the RTD is very slow (i.e., $\min(\sub{t}{v}, \sub{t}{i}) \gg \max(\sub{t}{s}, \sub{t}{n})$), the laser equilibrates almost instantly relative to the characteristic time scale of the RTD. In this case, we can adiabatically reduce the four-dimensional system state $(v,i,s,n)$ to the two-dimensional state $(v,i)$ by setting $\dot{s} = 0$ and $\dot{n} = 0$. With \cref{eq:system_s,eq:system_n}, this leads to
	\begin{align}
		0 &= (n - 1) s + \frac{\sub{\gamma}{m}}{\sub{\gamma}{t}} (n_0 + n) \,, \\
		0 &= \hat{j} - n (1 + s) \,,
	\end{align}
	where we define the combined bias current $\hat{j}(t) = j + \eta i(t)$ for brevity. Solving this system of equations, we arrive at an approximation of the delayed term $s(t-\tau)$ by the nonlinear function
	\begin{align} 
		s(t) &= \frac{1}{2}
		\Bigg(
		\nu - 1 + \hat{j}(t) \label{app:eq:s_limit_solution}\\
		&\hspace{19pt}+ \sqrt{(1 + \nu)^2 + 2 \bigg(\frac{\sub{\gamma}{m}}{\sub{\gamma}{t}} (n_0 + 2) - 1\bigg) \hat{j}(t) + \hat{j}(t)^2}
		\Bigg) \notag
	\end{align}
	with the shorthand $\nu = \frac{\sub{\gamma}{m}}{\sub{\gamma}{t}} n_0$. The system equations are thus
	\begin{align}
		\sub{t}{v} \dot{v}
		&= i - f(v) - \kappa s(t - \tau) \label{eq:system_noisefree_v_slowRTD}\,,\\
		\sub{t}{i} \dot{i}
		&= v_0 - v - r i \label{eq:system_noisefree_i_slowRTD}\,.
	\end{align}
	Importantly, the simplified system, \cref{eq:system_noisefree_v_slowRTD,eq:system_noisefree_i_slowRTD}, is closely related to the prototypical delayed FitzHugh--Nagumo model of the RTD in Ref.~\cite{RomeiraAFBJ2016} except for the more realistic current-voltage characteristic $f(v)$ and the non-Pyragas feedback term~\cite{Pyragas1992,SchoellS2007}.

	\subsection{\label{sec:methods_numerical_methods}Numerical methods}
	
	\begin{figure}[t!]
		\centering
		\includegraphics[width=\linewidth]{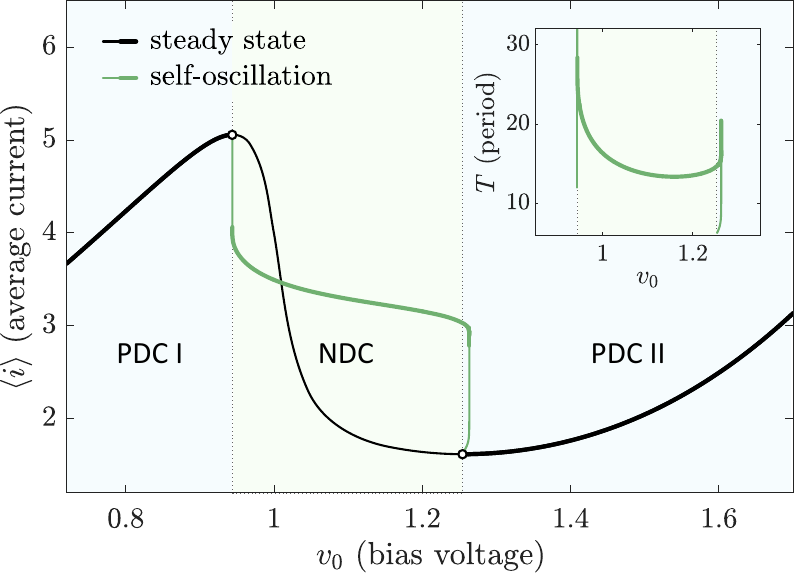}
		\caption{ Bifurcation diagram of the RTD system~\mbox{\eqref{eq:system_v}-\eqref{eq:system_i}} in the bias voltage $v_0$ without feedback ($\kappa = 0$). The thick lines represent stable solutions, the thin lines unstable solutions. The self-oscillation branch (green) attaches to the steady state (black) through two subcritical Andronov-Hopf bifurcations (white circles). Note that the slope of the canard explosions near the two AH bifurcations is almost vertical. In the inset, the period $T$ of the self-oscillation branch is shown.}
		\label{fig:2}
	\end{figure}

	In the theoretical analysis of the RTD-LD system, we employ both time simulations and path-continuation methods. For the former, a semi-implicit numerical scheme is employed to solve 
	DDEs \mbox{\eqref{eq:system_v}-\eqref{eq:system_n}}, see \cref{app:sec:numerical_solution} for details. 
	
	The path continuation for the bifurcation analysis of the DDE system \mbox{\eqref{eq:system_v}-\eqref{eq:system_n}} relies on the \textsc{Matlab} package \textsc{DDE-BifTool}~\cite{EngelborghsLR2002}. The code for the bifurcation analysis and corresponding visualizations of the bifurcation diagrams are freely available~\cite{ZenodoMMGJ2024}. In the following, we use the integrated intensity over one period~$T$,
	\begin{equation}
		\average{x} = \frac{1}{T} \INT{0}{T}{t} x(t)\,,
	\end{equation}
	as measure for periodic solutions.
	
	The specific parameter values we fixed in the following for concreteness of the model are listed in \cref{app:sec:typical_parameters}. Unless mentioned otherwise, the delay time is fixed at $\tau = 20$, corresponding to different \emph{physical} delays $\tau t_\mathrm{c}$, depending on the characteristic time scale $t_\mathrm{c}$.
	Furthermore, we classify the characteristic time scale, defined as the RTD self-oscillation (tank) frequency $\sub{t}{c} = \sqrt{LC}$ in \cref{app:eq:characteristic_timescale}, into two regimes. For the fast RTD, we select $t_{\mathrm{c},\mathrm{fast}} \approx 15.9 \,\si{\pico\second}$ with capacitance $C = 2\,\si{\femto\farad}$ and inductance $L = 126\,\si{\nano\henry}$. In contrast, for the slow RTD, we set $t_{\mathrm{c},\mathrm{slow}} \approx 15.9 \,\si{\nano\second}$ with $C = 2\,\si{\pico\farad}$ and inductance $L = 126\,\si{\micro\henry}$. The slow RTD thus operates a thousand times slower than the fast RTD so that the slow-RTD approximation applies.
	
	\section{\label{sec:results}Results and discussion}
	
	\subsection{\label{sec:results_noFeedback}No feedback}
	
	To understand the influence of the delayed feedback on the system in question, let us first review how the RTD system~\mbox{\eqref{eq:system_v}-\eqref{eq:system_i}} operates without feedback by setting $\kappa = 0$. A comprehensive bifurcation analysis of this case has been performed in Ref.~\cite{OPPFRJ2021}.
	The bifurcation diagram in~\cref{fig:2} shows that the steady state (black) indeed resembles $f(v)$ closely. On either side of the NDC region, the steady state loses its stability in a subcritical Andronov-Hopf (AH) bifurcation (white circles), so that the emerging periodic solution (green) coexists with the steady state in a small region of bistability within PDC~II. The sudden increase in the amplitude of the limit cycle around the bifurcation points indicates a \emph{canard explosion}. First discovered in 1981~\cite{BenoitCDD1981}, the canard is a rapid transition from small-amplitude to large-amplitude limit cycles by varying a control parameter in an exponentially narrow range~\cite{Izhikevich2007}. Canards are associated with excitable systems such as neurons~\cite{DesrochesKR2013,RomeiraFJ2017,ZhaoG2017,OPPFRJ2021,GirierDR2023} but also orgasms~\cite{BlyussK2023}, where they can induce quasi-thresholds.
	
	The subcriticality of the two AH bifurcations differs from the FHN model, where the AH bifurcations are supercritical. Since the slope of the canard explosion near the two AH bifurcations is almost vertical, in particular at the border of the PDC~I region, the region of bistability cannot be visualized because it is smaller than the numerical accuracy of the branch points. As pointed out in Refs.~\cite{WallisT1994,OPPFRJ2021}, the reason for the subcriticality lies in the current-voltage characteristic~$f(V)$ in \cref{eq:current-voltage}. Our system exhibits richer dynamics than the FHN model in part due to this more intricate current-voltage characteristic. Typical canard trajectories along the slow manifold can indeed be observed, albeit unstable due to the subcriticality of the AH bifurcations, in the video of solutions along the self-oscillation branch from \cref{fig:2} in the Supplemental Material~\cite{ZenodoMMGJ2024}.
	
	Given that the periodic solution does not rely on external perturbations and maintains a characteristic period $T$ (cf.~the inset in \cref{fig:2}) that is largely independent of the delay $\tau$, this solution is called \emph{self-oscillation}. \Cref{fig:2} shall serve as reference point for our subsequent analysis to comprehend how time-delayed feedback alters this picture.

	\subsection{\label{sec:results_slowRTD}Regime of the slow RTD}
	
\begin{figure}[h!]
		\centering
		\includegraphics[width=\linewidth]{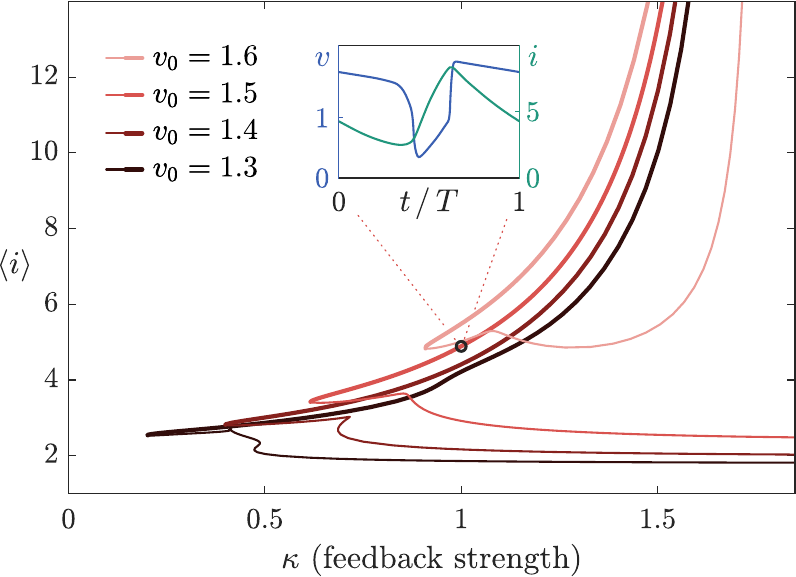}
		\caption{Bifurcation diagram of the the slow-RTD regime as given by Eqs.~\eqref{eq:system_noisefree_v_slowRTD}-\eqref{eq:system_noisefree_i_slowRTD} in the feedback strength $\kappa$ for four different values of $v_0$. The inset shows a typical solution profile.}
		\label{fig:3}
	\end{figure}

	Let us now turn to the effect of time-delayed feedback on the RTD-LD dynamics in the slow-RTD approximation described in \cref{sec:methods_slow-RTD_approximation} by setting the characteristic time scale to $t_{\mathrm{c},\mathrm{slow}}$. Interestingly, a continuation of the corresponding DDE system~\eqref{eq:system_noisefree_v_slowRTD}-\eqref{eq:system_noisefree_i_slowRTD} in the feedback strength $\kappa$ for different values of $v_0$ as shown in \cref{fig:3} reveals that a new kind of solution emerges, see the inset for a typical time trace. This solution exists for a range of bias voltages $v_0$ in the PDC II region if the feedback is sufficiently strong.
	Note that the stable parts of all four branches diverge to infinity at $\kappa \approx 1.5$. The reason for this divergence is a feedback catastrophe, similar to that of a microphone held too close to a coupled loudspeaker: If the feedback is strong enough, each round trip in the circuit injects more and more energy into the circuit.

	\begin{figure*}[tbhp]
		\centering
		\includegraphics[width=\linewidth]{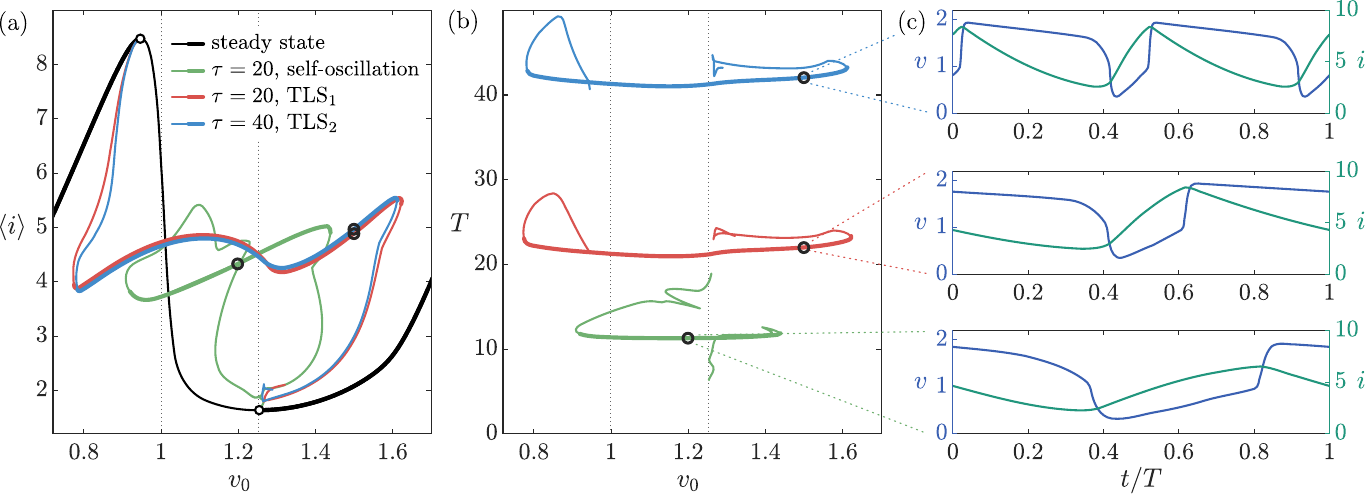}
		\caption{(a) Continuation in the bias voltage $v_0$ at $\kappa = 1$ in the slow-RTD regime. The steady state (black) loses its stability through two Andronov-Hopf bifurcations (white circles). (b) Period $T$ of the corresponding periodic branches. (c) Time traces at $v_0 = 1.2$ for self-oscillation (green) and $v_0 = 1.5$ for the $\mathrm{TLS}_1$ (red) and $\mathrm{TLS}_2$ (blue). The position of the time traces in (a) and (b) is indicated by black circles. Vertical dashed lines separate the differential conductance regions.}
		\label{fig:4}
	\end{figure*}
	To understand this new solution, we now fix the feedback strength at $\kappa = 1$ and follow the solutions of system~\eqref{eq:system_noisefree_v_slowRTD}-\eqref{eq:system_noisefree_i_slowRTD} in the bias voltage $v_0$, see~\cref{fig:4}~(a).
	First, we observe that the steady-state branch (black) has shifted relative to the case $\kappa=0$ (cf.~\cref{fig:2}) because the non-Pyragas feedback term, $\kappa s(t - \tau)$, injects energy and thus raises the steady state to a higher average current~$\average{i}$, in contrast to the non-invasive Pyragas feedback, $\kappa \big(s(t-\tau) - s(t)\big)$, used in Ref.~\cite{RomeiraAFBJ2016}. Second, the self-oscillation branch (green) has twisted into a loop and extends from the NDC into the PDC~II region, separated by dashed vertical lines. 
	However, the interesting difference to the case $\kappa = 0$ is a new solution type presented in~\cref{fig:3}: the red and the blue branch, labeled $\mathrm{TLS}_1$ and $\mathrm{TLS}_2$, which are one- and two-pulse \emph{temporal localized states} (TLSs), respectively. They emerge from the steady state in an AH bifurcation and remain stable within the NDC and parts of the PDC I and II regions, see panel~(c) for typical profiles. Note that we chose the measure in the bifurcation diagram such that the two branches $\mathrm{TLS}_1$ at $\tau = 20$ and $\mathrm{TLS}_2$ at $\tau = 40$ have a similar shape. A~two-pulse state $\mathrm{TLS}_2$ does not exist for $\tau = 20$ because the domain is too small, yet there is a one-pulse state $\mathrm{TLS}_1$ for $\tau = 40$. For any delay $\tau$, there coexist as many $\mathrm{TLS}_n$ solutions as pulses can fit into the temporal domain. In fact, self-oscillation is the limit of the $\mathrm{TLS}_n$ where the entire domain is filled with tightly packed single-pulse TLSs, cf.~Ref.~\cite{RomeiraAFBJ2016}. We chose to double $\tau$ for $\mathrm{TLS}_2$ because $\tau = 20$ would be too small to fit two pulses. 
	The fact that $\mathrm{TLS}_1$ and $\mathrm{TLS}_2$ connect to the left AH bifurcation at the boundary of the PDC I region is actually a finite-size effect of the relatively small temporal domains $\tau = 20$ and $\tau = 40$, which fit one and two pulses, respectively.
	
	The nature of the TLSs becomes clearer when considering the period $T$ in \cref{fig:4}~(b) and corresponding time traces of one period at exemplary bias voltages (black circles) in panel~(c). The stable part of the self-oscillation solution has a period of $T \approx 11$ and fills the entire domain with oscillations, as seen in the bottom plot of panel~(c). In contrast, the period of the $\mathrm{TLS}_1$, which is stable around $T \approx 22$, depends on the delay. In fact, its period $T = \tau + \delta$, is slightly larger by a drift $\delta$ than the delay $\tau$ due to causality \cite{YanchukG2022}. In the context of excitability, this drift corresponds to the latency between when a perturbation triggers the system and the ensuing excited orbit. Similarly, $\mathrm{TLS}_2$ has double the period, $T = 2\tau + \delta_2 \approx 42$, with some other drift $\delta_2$, as we can see from panel~(b), cf.~Ref.~\cite{YanchukRSW2019}.
	
	Let us emphasize that there are two significant features of the TLSs presented in \cref{fig:4} with respect to memory. First, the TLSs coexist with the steady state. In conjunction with the excitability of the RTD, this bistability means that the stable steady state can be perturbed in the PDC II region, triggering a pulse. The TLSs are stabilized by the feedback, without which the pulse would be a single excursion through the phase space and back to the steady state. However, the feedback is strong enough to sustain the pulse on its next round trip, emulating a series of neurons propagating the pulse.
	
\begin{figure}[b]
		\centering
		\includegraphics[width=\linewidth]{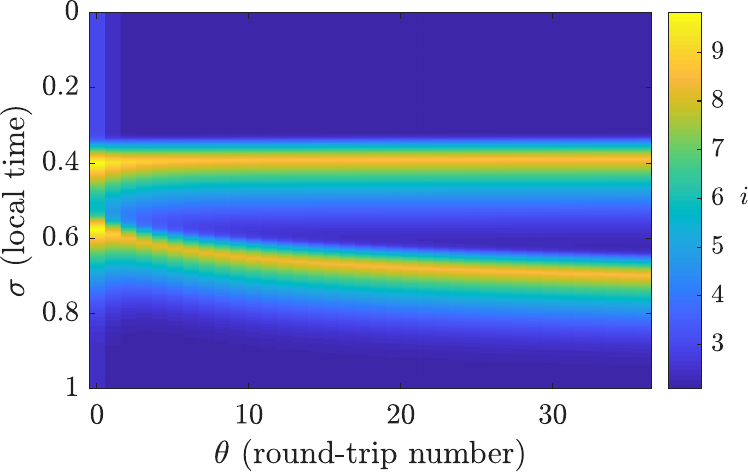}
		\caption{Two-time diagram of a $\mathrm{TLS}_2$ in the slow-RTD regime. The pulses repel each other from an initial distance $d_0 = 0.17 \tau \approx 0.166 T$ in local time $\sigma$ with feedback strength $\kappa = 1$ and delay $\tau = 80$ at bias voltage $v_0 = 1.5$. The local time $\sigma$ is relative to the period $T = \tau + \delta \approx 81.8$ with the drift~$\delta$.}
		\label{fig:5}
	\end{figure}

	Second, the $\mathrm{TLS}_2$ solution in which the two TLSs are equidistant is stable. To explain this stability, the question is how two pulses racing around the RTD-LD circuit affect each other. An extremely useful method for answering this question is the \emph{two-time representation}~\cite{ArecchiGLM1992,GiacomelliP1996,YanchukG2022}. The motivation for the two-time representation is to highlight dynamics on vastly different time scales, the period $T = \tau + \delta$ on the one hand and the dynamics over many round trips of the pulses within the circuit on the other. This representation is achieved by parameterizing time as $t = (\theta + \sigma) T$ via the number of round trips $\theta = \floor{t/T}$ and the local time  $\sigma = t/T \mod 1$ within the most recent round trip, and then plotting the time trace in the $(\theta,\sigma)$-plane. The current $i$ of the time trace is best suited to represent the pulse dynamics, as it is a slow variable. \cref{fig:5} presents a two-time representation of a time simulation, initialized with two non-equidistant pulses (initial distance $d_0 = 0.17 \tau$) within the temporal domain $[-\tau,0)$. Note that if the pulses are initiated even closer to each other, time simulations show that both pulses die immediately and the system jumps to the steady state.

	The key observation here is that, while a single pulse would move horizontally in the two-time diagram (since the drift $\delta$ has been accounted for in the definition of $\sigma$ and $\theta$), the two pulses \emph{repel} each other. The non-reciprocal repulsive interaction is most pronounced for the second pulse starting at about $\sigma = 0.6$, but since the two-time parameterization transforms the time $t$ into helical boundary conditions, the second pulse also interacts with the first. 

	\begin{figure*}[tbhp]
		\centering
		\includegraphics[width=\linewidth]{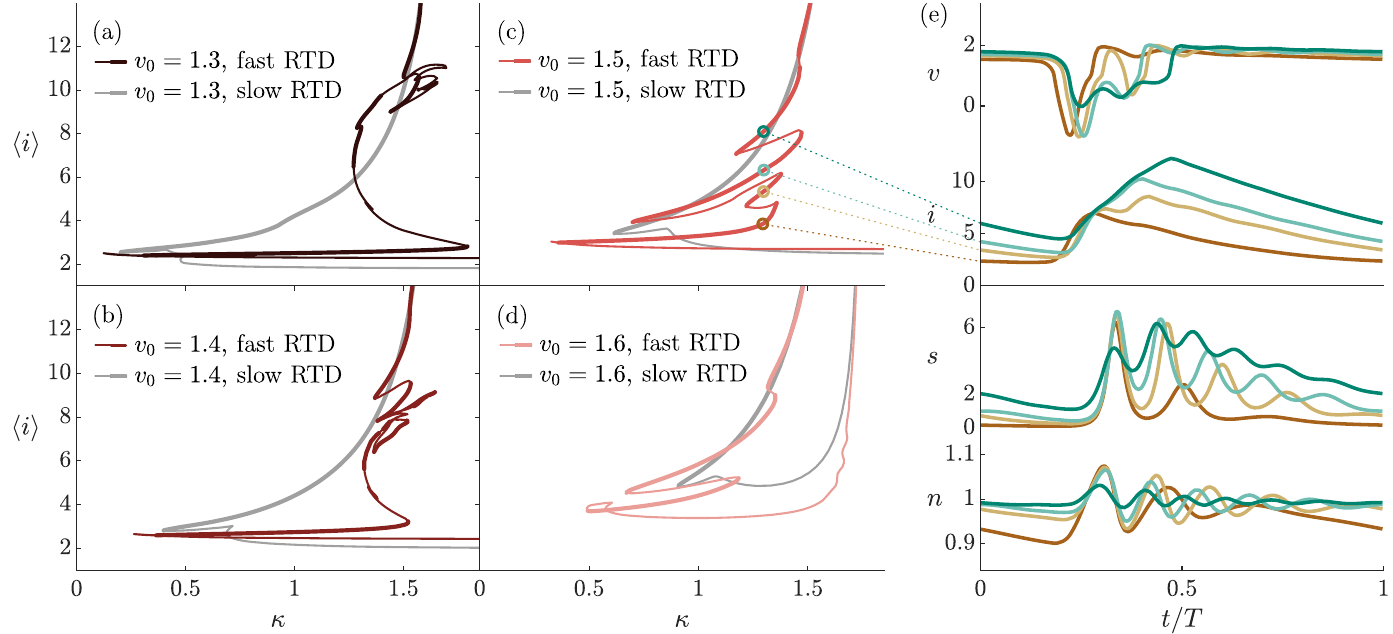}
		\caption{(a)-(d) Continuation of $\mathrm{TLS}_1$ in the feedback strength $\kappa$ for different bias voltages $v_0$ in the fast (shades of red) and slow (gray, cf.~\cref{fig:3}) RTD regimes. (e) Time traces of the multistable $\mathrm{TLS}$ solution at $v_0 = 1.5$ and $\kappa = 1.3$.}
		\label{fig:6}
	\end{figure*}
	It is well known that TLSs interact via tail overlap~\cite{VladimirovMSF2002,TlidiVM2003}, and in our case, as we operate in the slow-RTD regime, the dynamics are essentially controlled by the current $i$, which is the slowest variable. An interaction law between TLSs could be derived following, e.g., Ref.~\cite{MunsbergJG2020}. However, intuition about the excitability mechanism is sufficient to understand the nature of the repulsive forces. When the feedback arrives at the RTD to trigger the second TLS, the system has not quite reached the steady state yet since the memory of the first TLS is preserved in its tail. In particular, the value of the current $i$ has not fully recovered. Therefore, the excitability threshold for the second pulse is slightly higher and, although the feedback is strong enough to cross this threshold, the pulse is slightly delayed. Thus, the distance between the two TLSs increases, resulting in an effectively repulsive interaction. We note that since the interaction is mediated by a slow variable recovery, the pulses interact almost exclusively forward in time. Such non-reciprocal interactions are typical for time-delayed systems, e.g., Ref.~\cite{MunsbergJG2020}, where non-reciprocal interaction between TLSs in a type I excitable system based on the delayed Adler model is discussed.
	
	The results of this subsection assuming a slow RTD -- the stable TLS branch coexisting with the excitable, stable steady state, repulsive TLS interactions, and even the winged shape of the TLS branches -- are qualitatively strikingly similar to the simple delayed FHN neuron model of an RTD-LD studied in Ref.~\cite{RomeiraAFBJ2016}. There it was shown that the $n$ pulses in a $\mathrm{TLS}_n$ can be manipulated independently and may thus serve as memory. We can conclude that the approximation of using the FHN model in Ref.~\cite{RomeiraAFBJ2016} is justified to qualitatively reproduce the pulse dynamics in the adiabatic limit of a slow RTD. Moreover, TLSs are suitable to act as memory in the RTD-LD because of their robustness to perturbations, called self-healing, and their repulsive interaction, which allows information to be stored over long periods of time.

	\subsection{\label{sec:results_fastRTD}Regime of the fast RTD}

	Now, we consider the scenario where the RTD and the laser evolve on similar time scales by setting the characteristic time scale of the RTD to $t_{\mathrm{c},\mathrm{fast}} = t_{\mathrm{c},\mathrm{slow}} / 1000$, cf.~\cref{sec:methods_numerical_methods}. A fast RTD means that the adiabatic approximation from \cref{sec:methods_slow-RTD_approximation} used in \cref{sec:results_slowRTD} is no longer justified and we have to consider the complete RTD-LD system~\mbox{\eqref{eq:system_v}-\eqref{eq:system_n}}. In this scenario, we anticipate the time scale of the laser relaxation oscillations to interact with the RTD spiking period.

	\begin{figure*}[tbhp]
		\centering
		\includegraphics[width=\linewidth]{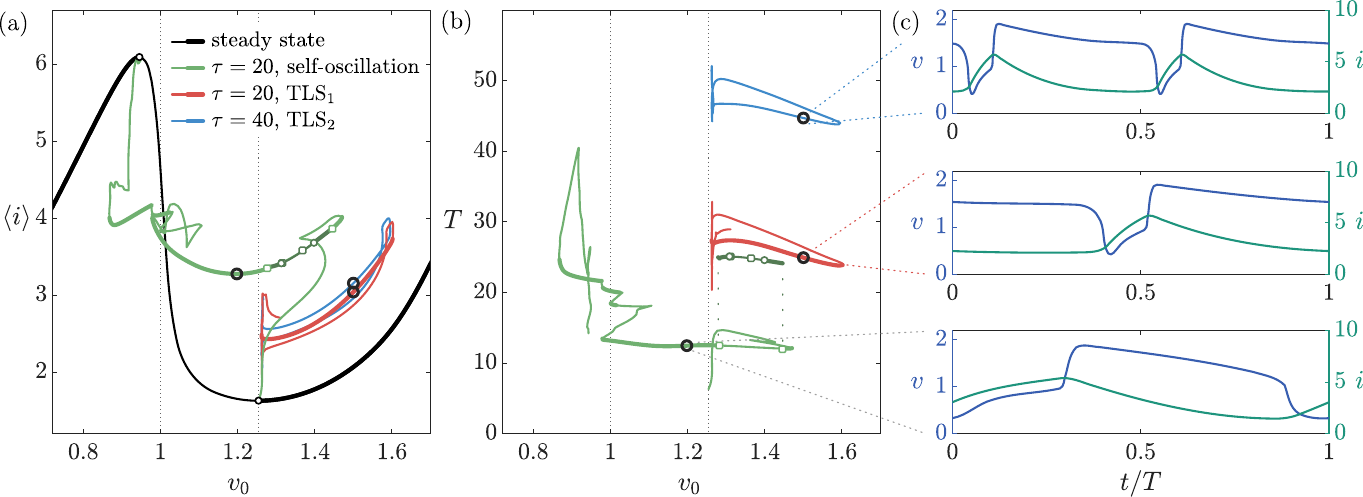}
		\caption{(a) Continuation in the bias voltage $v_0$ at $\kappa = 0.5$ in the  fast-RTD regime. The steady state (black) loses its stability through two AH bifurcations (white circles). The green, red, and blue branches correspond to the periodic self-oscillation, $\mathrm{TLS}_1$, and $\mathrm{TLS}_2$, respectively. (b) Period $T$ of the periodic branches. The dark green branch arises through a period doubling (white squares) of the self-oscillation. (c) Exemplary time traces in voltage $v$ and current $i$ of the solutions at $v_0 = 1.2$ for self-oscillation and $v_0 = 1.5$ for the $\mathrm{TLS}_1$ and $\mathrm{TLS}_2$. The position of the time traces in (a) and (b) is indicated by black circles.} 
		\label{fig:7}
	\end{figure*}
	\Cref{fig:6} shows a continuation in the feedback strength~$\kappa$, along with the branches from \cref{fig:3} for a slow RTD in gray. Panels (a) to (d) are slices at different bias voltages $v_0 \in \{1.3, 1.4, 1.5, 1.6\}$ in the PDC~II region.
	As in \cref{fig:3}, the TLSs exist and are stable above a threshold in $\kappa$ that is similar to the slow-RTD scenario for $v_0 \in \{1.3, 1.4\}$, but the onset is lower for $v_0 \in \{1.5, 1.6\}$. Time simulations confirm that the TLSs arise due to excitability of the steady state, just as in the slow-RTD regime. Furthermore, the large-scale behavior of the branches is similar with respect to the resonance catastrophe, where the branches diverge to infinity at $\kappa = 1.5$. However, the laser dynamics lead to much more intricate branches, which can be attributed to relaxation oscillations of the laser.
	
	The most striking difference between the fast-RTD and slow-RTD regime is that the branch is monostable from $\kappa = 0.3$ to $\kappa = 0.7$ but distorts into a number of multistable patches around $\kappa = 1.5$, whereas the slow RTD branch has a single monotonous stable patch. Multistability has been reported in excitable time-delayed systems before, e.g., the multistability of pulse numbers in the laser cavity for the delayed Yamada model~\cite{TerrienEtAl2020, TerrienPKBB2021}. To illustrate the significance of the multistable periodic solutions, panel~(e) compares the time traces of four exemplary $\mathrm{TLS}$~solutions of different energy coexisting at the same bias voltage $v_0 = 1.5$ and feedback strength $\kappa = 1.3$. The multistability of the TLSs is particularly interesting because it could enable nonbinary encoding. Yet the four solutions have slightly different periods, $T \in \{22.86, 22.34, 22.13, 22.08\}$, which implies that they move at different speed around the circuit, rendering nonbinary encoding unstable.
	
	A continuation in the bias voltage $v_0$ at fixed $\kappa = 0.5$, as shown in~\cref{fig:7}, reveals crucial differences to the analogous \cref{fig:4} for a slow RTD. In panel~(a), the self-oscillation branch (green) connects to the steady state (black) at two AH bifurcations (white circles) and is much more twisted, including an intricate multistable region near $v_0 = 1$. Further, the $\mathrm{TLS}_1$ (red) and $\mathrm{TLS}_2$ (blue) branches attach to the canard explosion of the self-oscillation branch. This reordering of where the branches attach can be explained by shorter pulse widths, which reduce finite domain size effects. Furthermore, around $v_0 = 1.3$ and $v_0 = 1.5$, the self-oscillation branch loses stability in two period-doubling (PD) bifurcations (white squares), between which the emerging branch in dark green (whose period is indeed twice as large, see panel~(b)) itself loses stability in a pair of AH bifurcations, followed by another pair of PD bifurcations. However, the spectrum of Floquet multipliers shows that the solution at about $v_0 > 1.26$ is only marginally stable; a manual continuation with time simulations jumps to $\mathrm{TLS}_1$ or the steady state branch. 	
	Note that the period $T$ of the self-oscillation in panel~(b) is generally similar to the solution without feedback presented in~\cref{fig:2}, i.e., convex with unstable legs downward near the boundary of the NDC region. We do not see this shape in the analogous \cref{fig:4}, again due to finite-size effects.
	The striking observation in this figure, however, is that while the $\mathrm{TLS}_1$ branch is stable for a wide range of $v_0$, the $\mathrm{TLS}_2$ branch is entirely unstable. Panel~(b) illustrates that, while the self-oscillation branch is independent of the delay $T$, the $\mathrm{TLS}$ branches have a period $T = \tau + \delta$ that changes with the delay through a shorter or longer resting time while the pulse remains the same. The exemplary time traces of the periodic solutions at $v_0 = 1.5$ in panel~(c) demonstrate that the TLSs are indeed localized perturbations to a background resting state, but their qualitative shape remains mostly the same, whereas the self-oscillation fills the whole temporal domain.

	\begin{figure}[tbhp]
		\centering
		\includegraphics[width=\linewidth]{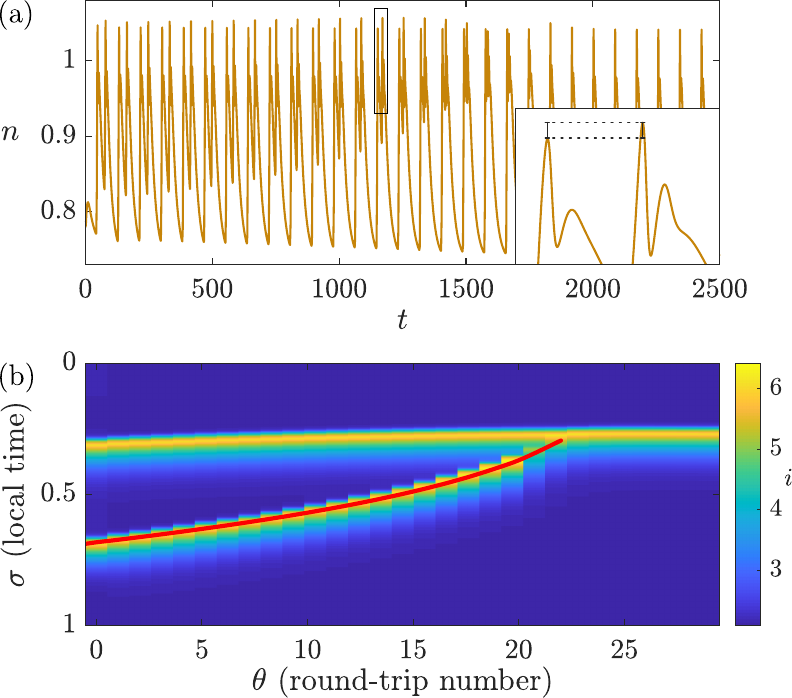}
		\caption{(a) Time series of the carrier number $n$ of the $\mathrm{TLS}_2$ at $v_0 = 1.5$, $\kappa = 1$, and $\tau = 80$ in the fast-RTD regime. The inset of one period shows that the second pulse is larger than the first. (b) The corresponding two-time diagram of two TLSs generated from an initial condition with a pair of spikes spaced a distance~$d_0 = 0.4 \tau \approx 0.376 T$ apart in local time~$\sigma$. The folding factor is $T = \tau + \delta_2 \approx 85.09$. The fit (red line) of the position of the second pulse with fit parameter $(50.3 \pm 0.3) \cdot 10^{-3}$ corresponds to the solution of the equation of \mbox{motion \eqref{eq:TLS_equation_of_motion}}.}
		\label{fig:8}
	\end{figure}
	The question that remains is why the $\mathrm{TLS}_2$ branch is unstable. By initializing two non-equidistant pulses at a distance $d_0 = 0.4 \tau \approx 0.376 T$ in local time $\sigma$, we find in \cref{fig:8} (a) that the two respective pulses in the carrier number $n$ move closer and closer over time until they eventually merge at about $t = 1800$. The two-time representation of this transition for the current $i$ in panel~(b) reveals how the pulses interact as they race around the RTD-LD circuit. We conclude that the interaction of the pulses is \emph{attractive}. Consequently, the branch of the equidistant $\mathrm{TLS}_2$ solution in \cref{fig:6} is unstable due to this attractive, non-reciprocal interaction between the pulses.
	
	Notably in \cref{fig:8} (a), the second of the two spikes has a higher peak intensity, as shown in the inset. The difference in the peak height holds the key to understanding the mechanism by which the pulses attract each other. The second pulse occurs while the the carrier number $n$ of the first pulse has not yet fully relaxed. Consequently, the second pulse has more gain and is more intense than the first. Since the higher intensity leads to stronger feedback on the voltage $v$, the excitable response is slightly accelerated and thus, the second pulse catches up with the first. This mechanism occurs because the slow variable that governs the pulse interaction is the carrier number $n$, which effectively \emph{decreases} the excitability threshold in the feedback strength of a pulse. For the slow RTD, on the contrary, the incomplete recovery of the current $i$ \emph{increases} the excitability threshold. Our findings agree with repulsive and attractive interactions reported and explained for other excitable time-delayed systems, e.g., the delayed Yamada model~\cite{TerrienEtAl2018}.
	
	In \cref{app:sec:derivation_equations_of_motion_TLS}, we derive equations of motion for the pulse interaction from the simple ansatz that the interaction forces decay exponentially -- motivated by the exponential decay of the carrier number $n$ in the pulse tail with rate $\sub{t}{n}$ -- and act only forward in time because of causality. 
	The solution of these equations for the distance $d$ between two pulses with an initial distance $d_0$ in local time is
	\begin{align}
		d(\theta) = \frac{1}{2} + 2 \frac{\sub{t}{n}}{T} \arctanh\!\bigg( \ee^{B \theta} \tanh\!\bigg(\frac{T}{2\sub{t}{n}} \bigg[d_0 - \frac{1}{2}\bigg] \bigg) \bigg)\qquad\raisetag{4.22ex}
		\label{eq:TLS_equation_of_motion}
	\end{align}
	in terms of the round-trip number $\theta$, with a single free parameter $B = 2 A \exp\!\big(-\frac{T}{2\sub{t}{n}}\big) \frac{T}{\sub{t}{n}}$, where $A$ is the strength of the interaction force that contains the overlap integrals between the corresponding (adjoint) Goldstone modes and the tail of the interacting pulses. A fit based on \cref{eq:TLS_equation_of_motion}, represented by the red line in \cref{fig:8} (b), yields $B = (50.3 \pm 0.3) \cdot 10^{-3}$ and is in excellent agreement with the time simulation.

	\subsection{\label{sec:results_time-scale}Characteristic time scale}
	\begin{figure}[tbhp]
		\centering
		\includegraphics[width=\linewidth]{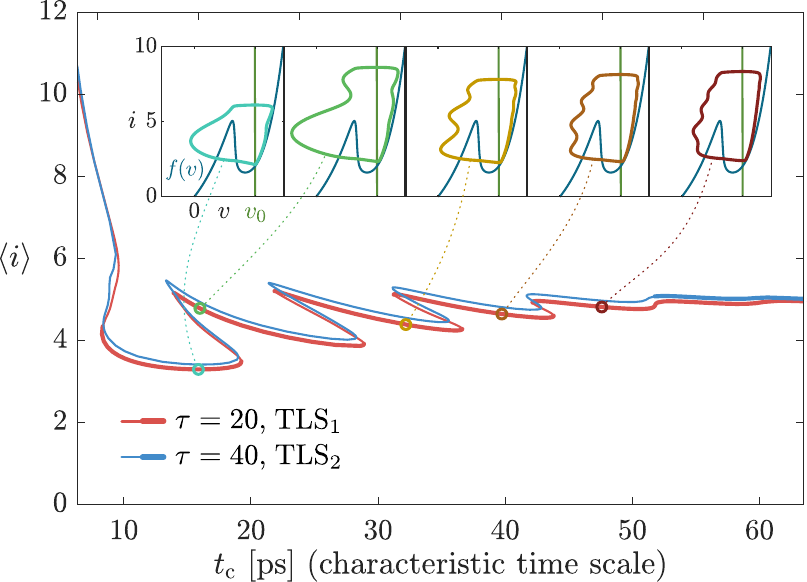}
		\caption{Continuation of the $\mathrm{TLS}_1$ and $\mathrm{TLS}_2$ solutions in the characteristic time scale $\sub{t}{c}$ at $v_0 = 1.5$ and $\kappa = 1$. The slow-RTD scenario is located at $\sub{t}{c} = 15.9 \,\si{\nano\second}$, the fast-RTD scenario at $\sub{t}{c} = 15.9 \,\si{\pico\second}$.}
		\label{fig:9}
	\end{figure}
	Having discussed the similarities and differences between the slow and the fast RTD, we finally consider the bifurcation diagram of the $\mathrm{TLS}_1$ and $\mathrm{TLS}_2$ solutions in the characteristic time scale $\sub{t}{c}$ shown in \cref{fig:9} to explain how these two regimes are related. In this diagram, the fast-RTD time scale corresponds to $\sub{t}{c} = 15.9 \,\si{\pico\second}$ and the slow RTD is located far to the right at $\sub{t}{c} = 15.9 \,\si{\nano\second}$ in the adiabatic limit (the branch remains relatively constant beyond~$\sub{t}{c} = 60 \,\si{\pico\second}$). There are two key features of the $\mathrm{TLS}$ solutions that we saw above to point out here.
	
	First, the $\mathrm{TLS}_2$ loses stability at around~$\sub{t}{c} = 52 \,\si{\pico\second}$ and remains unstable for smaller characteristic time scales. This is the point where the pulse interaction forces change. Apart from the difference in stability, the $\mathrm{TLS}_1$ and the $\mathrm{TLS}_2$ branch agree well even quantitatively.
	
	Second, both $\mathrm{TLS}$ branches exhibit slanted snaking in the fast-RTD domain, where the stability of the $\mathrm{TLS}_2$ breaks down. Notably, the stable patches of the snaking correspond to those of the continuation in~$\kappa$ in \cref{fig:6}. This horizontal snaking is another consequence of the complex laser dynamics. The insets in \cref{fig:9} of stable periodic solutions in the $(v,i)$-phase-space show that the stable patches evidently correspond to different numbers of oscillations in the tail, which appear as bumps in the phase-space trajectory, first one (cyan), then two, etc., up to five (brown). Note that the multistable solutions of the branches at $\kappa = 1$ and $v_0 = 1.5$ are exactly those two stable solutions from \cref{fig:6} (c) at $\kappa = 1$. Consequently, the multistability we discussed above has its origin in the snaking of the $\mathrm{TLS}_1$ branch in the characteristic time scale. However, with even stronger feedback, the snaking branches separate into islands, which further indicates that the dynamics of the RTD-LD are considerably convoluted.

	\section{\label{sec:conclusions}Conclusions}
	\vspace{-0.28cm}
	Our investigation of the RTD-LD system subjected to time-delayed feedback demonstrates with a realistic model derived from Ref.~\cite{OPHALHFRJ2022} that this optoelectronic circuit is indeed excitable and functions as an artificial neuron by generating temporally localized states (TLSs), which can serve as memory in neuromorphic computing.
	This analysis has far-reaching implications for actual devices and technology as it demonstrates under which conditions the RTD-LD device can be used for neuromorphic computation and memory buffers. We also explain the reason for these conditions, adding physical insight relevant even for a wider class of devices with slow-fast dynamics, namely that the fundamental limitations are due to the characteristic time scale of the RTD.
	The simpler delayed FitzHugh--Nagumo (FHN) model~\cite{RomeiraAFBJ2016} previously used to model the RTD-LD adequately describes the qualitative dynamics if the RTD is slow compared to the laser. In the slow-RTD limit, we find repulsive TLS interaction, which makes the memory stable. Yet our analysis has unveiled features and challenges previously unknown for the RTD-LD system, including the multistability of TLSs and attractive TLS interaction, if the RTD is fast, i.e., on a similar time scale as the laser. Although multistability could, in principle, enable nonbinary encoding, the periods of the coexisting TLSs types are not the same. Furthermore, the attractive interaction, which we could explain by the dynamics of the carrier number of the laser and describe analytically with excellent agreement, makes memory in the fast-RTD regime impossible, except for large delays. The importance of the laser dynamics is exemplified by the complex bifurcation scenarios with a fast RTD. We find a transition between the slow-RTD and fast-RTD regime through slanted snaking in the characteristic time scale of the RTD.
	
	Although the bifurcation analysis aims to provide a comprehensive understanding of how the more realistic model presented here agrees with or differs from simpler approximate models like the FHN model, numerical path continuation of the system proves very difficult. The self-oscillation branch serves to show the complexity introduced by the laser dynamics. We focus here on the parts relevant for understanding how to employ the RTD-LD as optoelectronic memory, but there is more to explore in terms of nonlinear dynamics in this system.
	
	In particular, it would be interesting to investigate which bifurcations connect the two regimes of the RTD time scale with different TLS interaction mechanism. 
	A~promising variation could be a smaller laser, which might yield higher speeds, but would require the increased noise to be addressed through Fokker-Planck equations as was already done for the FHN system~\cite{LindnerGONSG2004}. The prospect of experimental realization and eventually technological application, taking into account the theoretical findings in this paper, with all its caveats, stands as a promising research project to show that the RTD-LD might work in practice as an artificial neuron for neuromorphic computing.
	
	\section*{Data availability}
	The supporting code and supplemental videos for this article are openly available on Zenodo~\cite{ZenodoMMGJ2024}.
	
	\section*{Conflicts of interest}
	There are no conflicts of interest to declare.
	
	\begin{acknowledgments}
		We thank Bruno Romeira, Jos\'{e} Figueiredo, and Antonio Hurtado for helpful discussions about the RTD-LD dynamics.
		J.J.~acknowledges financial support from the European Commission through the H2020-FET-OPEN Project \textsc{ChipAI} under Grant Agreement 82884 and from the project KEFIR Grant PID2021-128910NB-I00 funded by MICIU/AEI/10.13039/501100011033 and by ERDF/EU.
		J.M.M.~thanks the Studienstiftung des deutschen Volkes for financial support. 
	\end{acknowledgments}
	
	\appendix
	
	\section{\label{app:sec:typical_parameters}System parameters}
	
	The system parameters are listed in \cref{app:tab:parameters} along with typical values in the numerical implementation. Most fixed values are consistent with those in Ref.~\cite{OPHALHFRJ2022}, but we have chosen a different value for $d$ to avoid a discontinuity in the current-voltage characteristic $f(v)$.

	\begin{table}[tbh]
		\caption{Overview of the model parameters and their typical values. Parameters that apply only to the non-dimensionalized system are highlighted in purple.}
		\label{app:tab:parameters}
		\begin{tabular}{|c l|r|}
			\hline
			& \textbf{RTD parameters} & \textbf{value}\\ \hline
			$a$    & $f(V)$ parameter & $-5.5 \cdot 10^{-5}\,\si{\ampere}$ \\
			$b$    & $f(V)$ parameter & $\SI{0.033}{\volt}$ \\
			$c$    & $f(V)$ parameter & $\SI{0.113}{\volt}$ \\
			$d$    & $f(V)$ parameter & $-3 \cdot 10^{-3}\,\si{\volt}$ \\
			$n_1$  & $f(V)$ parameter & $0.185$ \\
			$n_2$  & $f(V)$ parameter & $0.045$ \\
			$h$    & $f(V)$ parameter & $18 \cdot 10^{-5}\,\si{\ampere}$ \\
			$\kappa$ 						& optical feedback rate (LD$\to$RTD)	& varied \\
			\color{violet} $\kappa$ 		& 										& \color{violet} varied \\
			$R$    							& resistance 							& $\SI{10}{\ohm}$ \\
			\color{violet} $r$				& 										& \color{violet} $9.0 \cdot 10^{-4}$ \\
			$C$    							& capacitance 							& $2\,\{\si{\nano\farad}, \si{\femto\farad}\}$ \\
			$L$    							& inductance 							& $126\,\{\si{\micro\henry}, \si{\nano\henry}\}$ \\
			$V_0$  							& DC bias voltage 						& varied \\
			\color{violet} $v_0$ 			& 										& \color{violet} varied \\
			$\sigma$ 						& electrical noise amplitude 			& $0$ \\ \hline
			& \textbf{LD parameters} 				& \\ \hline
			$N_0$  							& transparency carrier number 			& $5 \cdot 10^5$ \\
			\color{violet} $n_0$ 			& 										& \color{violet} $2.5$ \\
			$\alpha$ 						& polarization factor 					& $0$ \\
			$\sub{\tau}{s}$ 				& photon lifetime 						& $5 \cdot 10^{-13}\,\si{\second}$ \\
			$\sub{\tau}{n}$ 				& carrier lifetime 						& $3.3 \cdot 10^{-10}\,\si{\second}$ \\
			$\sub{\gamma}{m}$ 				& Spont.\ em.\ into lasing mode 		& $10^7\,\si{\second^{-1}}$\\
			$\sub{\gamma}{l}$ 				& Spont.\ em.\ into leaky modes 		& $10^9\,\si{\second^{-1}}$ \\
			$\sub{\gamma}{nr}$ 				& non-radiative recombination		 	& $2 \cdot 10^9\,\si{\second^{-1}}$ \\
			$\eta$ 							& current injection efficiency (RTD$\to$LD) 	& $1$ \\
			\color{violet} $\eta$ 			& 										& \color{violet} $0.57$ \\
			$J$ 							& bias current in laser 				& $200\,\si{\micro\ampere}$ \\ 
			\color{violet} $j$ 				& 										& \color{violet} $-0.43$ \\
			$\tau$ 							& time delay of light coupling (LD$\to$RTD) 	& $\{0.32,0.63\}\,\si{\nano\second}$ \\
			\color{violet} $\tau$ 			& 										& \color{violet} $\{20,40\}$ \\ \hline
			& \textbf{derived parameters} 			& \\ \hline
			$\beta$ 						& spont. em. coupling $\beta = \sub{\gamma}{m} / (\sub{\gamma}{m} + \sub{\gamma}{l})$ 	& 0.01 \\ 
			$\sub{J}{th}$ 					& transcritical bifurcation value 		& $338\,\si{\micro\ampere}$ \\
			\color{violet} $\sub{j}{th}$ 	& 										& \color{violet} $1$ \\
			\color{violet} $\mu^2$ 			& \color{violet} stiffness 				& \color{violet} $1.96$ \\
			\color{violet} $\sub{t}{c}$ 	& \color{violet} characteristic time scale of RTD 	& \color{violet} $15.9\,\{\si{\nano\second}, \si{\pico\second}\}$ \\ \hline
			& \textbf{physical constants} 			& \\ \hline
			$\sub{q}{e}$					& elementary charge 					& $1.60\cdot10^{-19}\,\si{\coulomb}$ \\
			$\sub{k}{B}$ 					& Boltzmann constant 					& $1.38\cdot10^{-23}\,\frac{\si{\joule}}{\si{\kelvin}}$ \\
			$T$ 							& temperature 							& $\SI{300}{\kelvin}$ \\ \hline
		\end{tabular}
	\end{table}

	\section{\label{app:sec:model-derivation}Model derivation}
	
	In this section, we derive the RTD-LD model~\eqref{eq:system_v}-\eqref{eq:system_n} on the basis of a model presented in \cref{app:sec:model-derivation_physical-model} by performing a change of variables in \cref{app:sec:model-derivation_ito} and a non-dimensionalization in \cref{app:sec:model-derivation_nondim}.
	
	\subsection{\label{app:sec:model-derivation_physical-model}Physical model}
	The physical model for the voltage $V$, current $I$, electric field $E$, and carrier number $N$, adapted from Ref.~\cite{OPHALHFRJ2022} by adding a time-delayed feedback from the laser to the RTD, reads
	\begin{align}
		C \dot{V} &= I - f(V) - \kappa \abs{E(t-\tau)}^2 + \sigma \sub{\xi}{V}(t)\,, \label{app:eq:system_V}\\
		L \dot{I} &= V_0(t) - V - RI\,, \label{app:eq:system_I}\\
		\dot{E}   &= \frac{1 - \ii \alpha}{2} \bigg[ G  -  \frac{1}{\sub{\tau}{s}}\bigg] E
		\notag\\&\quad+ \sqrt{\frac{\sub{\gamma}{m} N}{2}} [\sub{\xi}{x}(t) + \ii \sub{\xi}{y}(t)]\,, \label{app:eq:system_S}\\
		\dot{N}   &= \frac{J + \eta I}{q} - \sub{\gamma}{t} N
		- G \abs{E}^2 \label{app:eq:system_N}
	\end{align}
	with the total decay rate $\sub{\gamma}{t} = \sub{\gamma}{l}+\sub{\gamma}{m}+\sub{\gamma}{nr}$, gain $G = \sub{\gamma}{m} (N - N_0)$, and the parameters listed in \cref{app:sec:typical_parameters}. The photon number $S$ is related to the complex electrical field $E = \sub{E}{x} + \ii \sub{E}{y}$ via $S = \abs{E}^2 =  \sub{E}{x}^2 + \sub{E}{y}^2$. As our model shall just consider the field intensity $S$, we set the Henry factor $\alpha = 0$ without loss of generality.
	Note that \cref{app:eq:system_V,app:eq:system_I,app:eq:system_S,app:eq:system_N} are stochastic delay-differential equations with uncorrelated Gaussian white noise $\sub{\xi}{V}$, $\sub{\xi}{x}$, and $\sub{\xi}{y}$ with zero mean $\E[\xi] = 0$ and auto-correlation $\average{\xi(t_1) \xi(t_2)} = \delta(t_1 - t_2)$, where $\sub{\xi}{V}$, $\sub{\xi}{x}$, and $\sub{\xi}{y}$ are mutually independent. For the scope of our analysis, however, we shall assume the noise in the RTD to be negligible by setting $\sigma = 0$, in accordance with time simulations.
	
	\subsection{\label{app:sec:model-derivation_ito}Change of variables}
	To arrive at a deterministic model, we aim to average the noise from the stochastic processes $\sub{\xi}{x}$ and $\sub{\xi}{y}$. It turns out to be convenient to use the photon number $S$ rather than the complex electric field $E$. For the transformation from $E$ to $S$, consider an \emph{It\^{o} drift-diffusion process} that satisfies the stochastic differential equation $\dif \vec{E}(t) = \vec{A} \dif t + \mat{B} \dif \vec{w}$, where we write the complex field as a vector $\vec{E} = (\sub{E}{x}, \sub{E}{y})^\mathrm{T}$ of real and imaginary parts, with $\vec{A} = \frac{1}{2} a \vec{E}$ and $\mat{B} = b \id$, and the \emph{Wiener process} (\emph{Brownian motion}) $\dif \vec{w} = (\sub{\xi}{x}(t + \dif t) - \sub{\xi}{x}(t), \sub{\xi}{y}(t + \dif t) - \sub{\xi}{y}(t))^\mathrm{T}$. In our model,
	\begin{align}
		a  &= G - \frac{1}{\sub{\tau}{s}}		,	\label{app:eq:noiseaverage_a}\\
		b  &= \sqrt{\sub{\gamma}{m} N / 2}		\,.	\label{app:eq:noiseaverage_b}
	\end{align}
	\emph{It\^{o}'s formula} states that for any transformation $g(t,x)$ (which is $C^2$) of an $n$-dimensional It\^{o} process $\dif \vec{X}(t) = \vec{A} \dif t + \mat{B} \dif \vec{w}$, the $k$th component of the It\^{o} process $\vec{Y}(t) = \vec{g}(t, \vec{X}(t))$ is described by (cf.~\citep[pp.~48]{Oksendal2003})
	\begin{equation}
		\dif Y_k = \pdif{g_k}{t} \dif t + \SUM{i=1}{n} \pdif{g_k}{x_i} \dif X_i + \frac{1}{2} \SUM{i,j=1}{n} \pdif{g_k}{x_i \partial x_j} \dif X_i \dif X_j\,.
	\end{equation}
	Consequently, for $g(t,x) = \abs{\vec{x}}^2$ and $S(t) = g(t, \vec{E}(t))$, we arrive after some calculation (using that $\dif X_i \dif X_j = \sum_{k=1}^{n} B_{ik} B_{jk} \dif t$ in this expansion to order $\dif t$ because $\dif w_i = \mathcal{O}(\sqrt{\dif t})$ and $\dif w_i \dif w_j = \delta_{ij} \dif t$) at
	\begin{equation}
		\dif S = a S \dif t + 2 b^2 \dif t + 2 b (\sub{E}{x} \dif \sub{w}{x} + \sub{E}{y} \dif \sub{w}{y})\,.
	\end{equation}
	Introducing the phase $\phi = \mathrm{atan2}(\sub{E}{y}, \sub{E}{x})$ of the field $\vec{E}$, we write the last term as 
	\begin{equation}
		2 b \big(\sqrt{S} \cos(\phi) \dif \sub{w}{x} + \sqrt{S} \sin(\phi) \dif \sub{w}{y}\big)\,,
	\end{equation}
	which is well defined since the noise variance goes to zero once $S$ approaches zero. So we can define a new stochastic process
	\begin{equation}
		\dif \sub{w}{S} = \big( \cos(\phi) \dif \sub{w}{x} + \sin(\phi) \dif \sub{w}{y}\big)\,,
	\end{equation}
	where $\sub{\xi}{S}$ defined by $\dif \sub{w}{S} = \sub{\xi}{S}(t + \dif t) - \sub{\xi}{S}(t)$ is again a Wiener process since $\Var[\sub{\xi}{S}] = \cos^2(\phi) \Var[\sub{\xi}{x}] + \sin^2(\phi) \Var[\sub{\xi}{y}] = 1$.
	With these definitions, we have
	\begin{equation}
		\dif S = (a S + 2 b^2) \dif t + 2 b \sqrt{S} \dif \sub{w}{S}\,,
	\end{equation}
	and reinserting $a$ and $b$ from above,
	\begin{equation} \label{app:eq:system_S_noisy}
		\dot{S} = \bigg(
		G - \frac{1}{\sub{\tau}{s}}
		\bigg) S
		+ \sub{\gamma}{m} N
		+ \sqrt{2 \sub{\gamma}{m} N S} \sub{\xi}{S}(t) \,,
	\end{equation}
	with noise variance $2 \sub{\gamma}{m} N S$. Note that Ref.~\cite{OPHALHFRJ2022} is missing a factor $2$ in the variance.
	
	\subsection{\label{app:sec:model-derivation_nondim}Non-dimensionalization}
	
	We define the dimensionless time~$\tilde{t}$ and delay~$\tilde{\tau}$ through $t = \sub{t}{c} \tilde{t}$ and $\tau = \sub{t}{c} \tilde{\tau}$, respectively, with the characteristic time scale $\sub{t}{c}$. Similarly, $V = \sub{v}{c} v$, $V_0 = \sub{v}{c} v_0$, $I = \sub{i}{c} i$, $S = \sub{s}{c} s$, and $N = \sub{n}{c} n + N_0$ define the dimensionless system variables $(v,i,s,n)$ and the bias voltage $v_0$.
	We shall determine a natural selection of characteristic scales by calculating the steady states of the system or approximations thereof.
	
	First, let us consider the steady state of the RTD. We see from \cref{app:eq:system_V} that the fixed points without feedback ($\kappa = 0$) are determined by $I = f(V)$ and $V \approx V_0$ if the resistance~$R$ is small. The current-voltage characteristic $f(V)$ has a jump height of order $a$ around \mbox{$c - n_1 V = 0$}, i.e., at $V = c / n_1$. With the characteristic scale for voltage and current defined as $\sub{v}{c} = c / n_1$ and $\sub{i}{c} = \abs{a}$, the new function $\tilde{f}(v) = f(V / \sub{v}{c}) / \sub{i}{c}$ has a jump of order~$1$ around $v = 1$.
	
	Next, we shall find the steady states of the LD. Solving $\dot{S} = 0$ and $\dot{N} = 0$ with Eqs.~\eqref{app:eq:system_S_noisy} and \eqref{app:eq:system_N} in the limit of a large laser (where $\sub{\gamma}{m} / \sub{\gamma}{t} = 0$), ignoring the average noise $\sub{\gamma}{m}N$, and assuming the current injection $I$ to be constant here for simplicity, we get two solutions: the off-state
	\begin{align}
		\sub{N}{off} &\approx \frac{J + \eta I}{\sub{\gamma}{t} q} \,,\\
		\sub{S}{off} &\approx 0 \,,
	\end{align}
	in which the laser does not emit photons,
	and the on-state 
	\begin{align}
		\sub{N}{on} &\approx \sub{n}{c} (1 +  n_0)\,,\\
		\sub{S}{on} &\approx \frac{\sub{\tau}{s} \sub{j}{c}}{q} \bigg(j + \tilde{\eta} i - 1\bigg) ,
	\end{align}
	where $\sub{n}{c} = 1 / (\sub{\tau}{s} \sub{\gamma}{m})$ is the characteristic scale along with the dimensionless transparency carrier number defined by $N_0 = \sub{n}{c} n_0$.
	We set the characteristic bias current to  $\sub{j}{c} = q \sub{\gamma}{t} \sub{n}{c} = \frac{q \sub{\gamma}{t}}{\sub{\tau}{s} \sub{\gamma}{m}}$ such that $J = \sub{j}{c} (j + n_0)$ and let $\tilde{\eta} = \eta \sub{i}{c} / \sub{j}{c}$.
	A natural choice for the characteristic photon number is
	\begin{equation}
		\sub{s}{c} = \frac{\sub{\tau}{s} \sub{j}{c}}{q} = \frac{\sub{\gamma}{t}}{\sub{\gamma}{m}} = \frac{1}{\sub{\tau}{n}\sub{\gamma}{m}}\,.
	\end{equation}
	In summary, the characteristic scales of the system variables and parameters are

	\hspace{-20pt}
	\begin{minipage}{0.5\linewidth}
		\begin{align}
			\sub{v}{c}   &= c / n_1 \,,\label{app:eq:def_vc}\\
			\sub{i}{c}   &= \abs{a} \,,\label{app:eq:def_ic}\\
			\sub{s}{c}   &= 1 / (\sub{\tau}{n} \sub{\gamma}{m}) \,,\label{app:eq:def_sc}\\
			\sub{n}{c}   &= 1 / (\sub{\tau}{s} \sub{\gamma}{m})\,,\label{app:eq:def_nc}\\\notag
		\end{align}
	\end{minipage}%
	\begin{minipage}{0.5\linewidth}
		\begin{align}
			\sub{j}{c}   &= q \sub{\gamma}{t} \sub{n}{c} \,,\label{app:eq:def_jc}\\
			\sub{\kappa}{c} &= \sub{i}{c} / \sub{s}{c}\,,\label{app:eq:def_kappac}\\
			\sub{r}{c}     &= \sub{v}{c} / \sub{i}{c}\,,\label{app:eq:def_rc}\\
			\sub{\eta}{c}  &= \sub{j}{c} / \sub{i}{c}  \,,\label{app:eq:def_etac} \\\notag
		\end{align}
	\end{minipage}
	where we also define the dimensionless resistance $r = R / \sub{r}{c}$ and feedback strength $\tilde{\kappa} = \kappa / \sub{\kappa}{c}$.
	
	To complete the derivation, we determine a characteristic time scale for the RTD and each of the four system variables $(v,i,s,n)$. 
	We begin by inserting the definitions~\eqref{app:eq:def_vc} to \eqref{app:eq:def_etac} of the rescaled variables into the system equations~\eqref{app:eq:system_V}, \eqref{app:eq:system_I}, \eqref{app:eq:system_S_noisy}, and \eqref{app:eq:system_N},
	\begin{align}
		\frac{C \sub{v}{c}}{\sub{i}{c} \sub{t}{c}} \tdif{v}{\tilde{t}}
		&= i - \tilde{f}(v) - \tilde{\kappa} s(\tilde{t} - \tilde{\tau}) \,,\\
		\frac{L \sub{i}{c}}{\sub{v}{c} \sub{t}{c}} \tdif{i}{\tilde{t}}
		&= v_0 - v - r i \,,\\
		\frac{\sub{\tau}{s}}{\sub{t}{c}} \tdif{s}{\tilde{t}}
		&= (n - 1) s + \frac{\sub{\gamma}{m}}{\sub{\gamma}{t}} (n + n_0) \notag\\&\quad+ \sqrt{2  \sub{\tau}{s} \sub{\tau}{n} \sub{\gamma}{m} (n + n_0) s} \sub{\xi}{S}(t) \,,\\
		\frac{\sub{\tau}{n}}{\sub{t}{c}} \tdif{n}{\tilde{t}}
		&= j + \tilde{\eta} i - n (1 + s)\,.
	\end{align}
	Although noise can have a profound impact on the dynamics of a system \cite{LindnerGONSG2004}, the noise here appears to be negligible \emph{after} the change of variables because, for typical parameters, the value $\sqrt{2 \sub{\tau}{s} \sub{\tau}{n} \sub{\gamma}{m}} \approx 6 \cdot 10^{-8}$ is much smaller than $n$ and $s$, which are of order one. Furthermore, time simulations verify that neglecting the noise is justified. Alternatively, we could derive the Fokker-Planck equation to obtain differential equations for the expected value of the state vector ($v,i,s,n$), but this would complicate matters unnecessarily.
	Notably, the prefactor of the average noise is 
	\begin{equation}
		\frac{\sub{\gamma}{m}}{\sub{\gamma}{t}} = \beta \mathrm{QE}\,,
	\end{equation}
	where the \emph{spontaneous emission coupling factor}~\cite{RomeiraF2018},
	\begin{equation}
		\beta = \frac{\sub{\gamma}{m}}{\sub{\gamma}{r}}\,,
		\label{app:eq:spont_em_coupling_factor_definition}
	\end{equation}
	is nonzero if the laser is small and the quantum efficiency is defined as
	\begin{equation}
		\mathrm{QE} = \frac{\sub{\gamma}{r}}{\sub{\gamma}{r} + \sub{\gamma}{nr}}
		\label{app:eq:QE_definition}
	\end{equation}
	with the radiative decay rate $\sub{\gamma}{r} = \sub{\gamma}{m} + \sub{\gamma}{l}$ and the total decay rate $\sub{\gamma}{t} = \sub{\gamma}{r} + \sub{\gamma}{nr}$.
	
	By setting
	\begin{equation}
		\mu = \frac{C \sub{v}{c}}{\sub{i}{c} \sub{t}{c}} = \bigg(\frac{L \sub{i}{c}}{\sub{v}{c} \sub{t}{c}}\bigg)^{-1} ,
	\end{equation}
	and solving the condition
	\begin{equation}
		1 = \mu \mu^{-1} = \frac{L C}{\sub{t}{c}^2}
	\end{equation}
	for the characteristic time scale $\sub{t}{c}$ of the RTD, we obtain
	\begin{equation} \label{app:eq:characteristic_timescale}
		\sub{t}{c} = \sqrt{LC}\,.
	\end{equation}
	A natural time scale of the photon number is the photon lifetime $\sub{\tau}{s}$, while for the carrier number the time scale is the carrier lifetime $\sub{\tau}{n}$. Since only the relative time scale of the LD versus the RTD is important for the dynamics of the system, we define the characteristic time scales of the LD relative to $\sub{t}{c}$, i.e.,
	\begin{minipage}{0.5\linewidth}
		\begin{align}
			\sub{t}{v} &= \mu 							\label{app:eq:t_v}\,, \\
			\sub{t}{i} &= \mu^{-1} 						\label{app:eq:t_i}\,, \\\notag
		\end{align}
	\end{minipage}%
	\begin{minipage}{0.5\linewidth}
		\begin{align}
			\sub{t}{s} &= \sub{\tau}{s} / \sub{t}{c} 	\label{app:eq:t_s}\,, \\
			\sub{t}{n} &= \sub{\tau}{n} / \sub{t}{c} 	\label{app:eq:t_n}\,. \\\notag
		\end{align}
	\end{minipage}
	We thus arrive at the system equations
	\begin{align}
		\sub{t}{v} \tdif{v}{\tilde{t}}
		&= i - \tilde{f}(v) - \tilde{\kappa} s(\tilde{t} - \tilde{\tau}) \label{app:eq:system_noiseaveraged_v}\,,\\
		\sub{t}{i} \tdif{i}{\tilde{t}}
		&= v_0 - v - r i \label{app:eq:system_noiseaveraged_i}\,,\\
		\sub{t}{s} \tdif{s}{\tilde{t}}
		&= (n - 1) s + \frac{\sub{\gamma}{m}}{\sub{\gamma}{t}} (n + n_0) \label{app:eq:system_noiseaveraged_s}\,,\\
		\sub{t}{n} \tdif{n}{\tilde{t}}
		&= j + \tilde{\eta} i - n (1 + s) \label{app:eq:system_noiseaveraged_n}\,.
	\end{align}
	Outside of this section, we omit the tilde on $\tilde{f}$, $\tilde{t}$, $\tilde{\tau}$, and $\tilde{\eta}$ and take the dot to mean the derivative with respect to $\tilde{t}$, e.g., $\dot{v} = \dif v / \dif \tilde{t}$.

	\section{\label{app:sec:numerical_solution}Time simulation}
	
	To solve the RTD-LD system numerically, we use a semi-implicit method.
	The coupling between the RTD and the LD is directed in the sense that the LD does not influence the RTD instantly, which means that within each step, we can first solve the RTD and then the LD.
	While we choose time steps $t_k = k h$ with step size $h$ and $k \in \N$ for the variables $(i, v, s)$, the carrier number $n$ is calculated as split stepping at $t_{k+\frac{1}{2}} = (k + \frac{1}{2}) h$.
	
	We obtain the numerical scheme by integrating the system equations \eqref{eq:system_v}-\eqref{eq:system_n} over one time step and averaging the variables between time steps. For the RTD, the discretization leads to
	\begin{align}
		\sub{t}{v} (v_{k+1} - v_{k}) &= \frac{h}{2} (i_{k} + i_{k+1}) - \INT{t_k}{t_{k+1}}{t} f(v) \notag\\
		&\quad- \frac{h}{2} \kappa (s_{k - \bar{\tau}} + s_{k + 1 - \bar{\tau}})\,,\\
		\sub{t}{i} (i_{k+1} - i_{k}) &= h v_0 - \frac{h}{2} (v_{k} + v_{k+1}) - \frac{h}{2} r (i_{k} + i_{k+1}) \,,
	\end{align}
	where the number of delay time steps is $\bar{\tau} = \tau / h$, where $h$ is chosen such that $\bar{\tau}$ is an integer.
	
	To approximate the integral, a Taylor expansion of $v(t)$ around $t_k$ to first order in $t$ and subsequently of $f(v)$ around $v_k$ to first order in $h$,
	\begin{align}
		\INT{t_k}{t_{k+1}}{t} f(v) &= \INT{t_k}{t_{k+1}}{t} f\bigg(v_k + \frac{t}{h} (v_{k+1} - v_k) + \mathcal{O}(t^2)\bigg) \\
		&= \INT{t_k}{t_{k+1}}{t}\;\, \bigg[f(v_k) +  \frac{t + \mathcal{O}(t^2)}{h} (v_{k+1} - v_k) f'(v_k) \bigg] \notag\\
		&= h f(v_k) + \frac{h}{2} (v_{k+1} - v_{k})  f'(v_k) + \mathcal{O}(h^2)\,, \notag
	\end{align}
	yields
	\begin{align}
		\sub{t}{v} (v_{k+1} - v_{k}) &= \frac{h}{2} (i_{k} + i_{k+1}) \notag\\
		&\quad -  h f(v_k)  - \frac{h}{2} (v_{k+1} - v_{k})  f'(v_k)\\
		&\quad - \frac{h}{2} \kappa (s_{k - \bar{\tau}} + s_{k + 1 - \bar{\tau}})\,,\notag\\
		\sub{t}{i} (i_{k+1} - i_{k}) &= h v_0 - \frac{h}{2} (v_{k} + v_{k+1}) - \frac{h}{2} r (i_{k} + i_{k+1}) \,. \raisetag{6.7ex}
	\end{align}
	Collecting the terms of step $k+1$ on the left, we rewrite the equations as
	\begin{align}
		a_{11} v_{k+1} + a_{12} i_{k+1} &= b_1\,, \\
		a_{21} v_{k+1} + a_{22} i_{k+1} &= b_2\,,
	\end{align}
	so that the solution for the step $k+1$ of RTD is
	\begin{align}
		v_{k+1} &= \frac{a_{22} b_1 - a_{12} b_2}{a_{11} a_{22} - a_{12} a_{21}}\,, \label{app:eq:system_numerical_v}\\
		i_{k+1} &= \frac{a_{11} b_2 - a_{21} b_1}{a_{11} a_{22} - a_{12} a_{21}}\,, \label{app:eq:system_numerical_i}
	\end{align}
	with
	\begin{equation}
		\begin{pmatrix}
			a_{11} & a_{12} \\ a_{21} & a_{22}
		\end{pmatrix}
		= \begin{pmatrix}
			\sub{t}{v} + \frac{h}{2} f'(v_k)  & -\frac{h}{2} \\ \frac{h}{2} & \sub{t}{i} + \frac{h}{2} r
		\end{pmatrix}
	\end{equation}
	and
	\begin{equation}
		\vectwo{b_1}{b_2} = \vectwo{\frac{h}{2} i_k  + \sub{t}{v} v_k - h \big[ f(v_k) - \frac{v_{k}}{2} f'(v_k) \big]}{h v_0 - \frac{h}{2} (v_{k} + r i_k) + \sub{t}{i} i_k}.
	\end{equation}
	The derivative $f'(v)$ necessary for the calculation is
	\begin{align}
		f'(v) &= - \sign(a) n_1 \frac{
			\log\!\Big(
			\frac{
				F^{+}(v) + 1
			}{
				F^{-}(v) + 1
			}
			\Big)
		}{
			d \Big(
			\frac{(c -  n_1 \sub{v}{c} v)^2}{d^2} + 1
			\Big)
		} \notag\\
		&\quad- \sign(a) n_1 \frac{q}{\sub{k}{B} T} 
		\frac{
			F^{+}(v) + F^{-}(v)
			\frac{
				F^{+}(v)  + 1
			}{
				F^{-}(v) + 1
			}
		}{
			F^{+}(v) + 1
		} \notag\\
		&\qquad \arctan\bigg(\frac{c - n_1 \sub{v}{c} v}{d} + \frac{\pi}{2}\bigg)\\
		&\quad+ \frac{h}{\abs{a}} n_2 \frac{q}{\sub{k}{B} T} \ee^{\frac{q}{\sub{k}{B} T} n_2 \sub{v}{c} v} \notag
	\end{align}
	with the abbreviation
	\begin{equation}
		F^{\pm}(v) = \ee^{\frac{q}{\sub{k}{B} T}(b - c \pm n_1 \sub{v}{c} v)}\,.
	\end{equation}

	For the LD, we arrive at
	\begin{align}
		\sub{t}{s} \big(s_{k+1} - s_{k}\big) &= \frac{h}{2} \big(n_{k+\frac{1}{2}} - 1\big) (s_{k+1} + s_{k}) \notag\\
		&\quad+ h \frac{\sub{\gamma}{m}}{\sub{\gamma}{t}} \big(n_{n + \frac{1}{2}} + n_0\big)\\
		&\quad + \sqrt{2 h \sub{\tau}{s} \sub{\tau}{n} \sub{\gamma}{m} (n_{k+\frac{1}{2}} + n_0) s_{k}} \xi_{\mathrm{S},k}(t), \notag
	\end{align}
	where $s_k$ in the optional noise term approximates $(s_{k+1} + s_{k})/2$,
	and by solving for $s_{k+1}$, we get
	\begin{align} \label{app:eq:system_numerical_s} \raisetag{2.5ex}
		s_{k+1} = \frac{
			\Big( \sub{t}{s} + \frac{h}{2} \big(n_{k+\frac{h}{2}} - 1\big)\Big) s_{k} + h \frac{\sub{\gamma}{m}}{\sub{\gamma}{t}} \big(n_{n + \frac{1}{2}} + n_0\big)
		}{
			\sub{t}{s} - \frac{h}{2} \big(n_{k+\frac{h}{2}} - 1\big)
		}. \hspace{28pt}
	\end{align}
	Similarly, we derive from 
	\begin{align}
		\sub{t}{n} \big(n_{k+\frac{3}{2}} - n_{k+\frac{1}{2}}\big) &= h (\eta i_{k+1} + j)\\
		&\quad- \frac{h}{2} \big(n_{k+\frac{1}{2}} +  n_{k+\frac{3}{2}}\big) (1 + s_{k+1})\notag
	\end{align}
	that
	\begin{align} \label{app:eq:system_numerical_n} \raisetag{4.6ex}
		n_{k+\frac{3}{2}} = 
		\frac{
			\Big(\sub{t}{n} - \frac{h}{2} (1 + s_{k+1}) \Big) n_{k+\frac{1}{2}} + h (\eta i_{k+1} + j) 
		}{
			\sub{t}{n} + \frac{h}{2} \big(1 + s_{k+1}\big)
		}.\hspace{30pt}
	\end{align}
	In summary, \cref{app:eq:system_numerical_v,app:eq:system_numerical_i,app:eq:system_numerical_s,app:eq:system_numerical_n} define the scheme for a time simulation of the RTD-LD model.
	
	\section{\label{app:sec:derivation_equations_of_motion_TLS}TLS equations of motion}
	
	In this section, we seek to model TLSs moving in the parameter space $(\theta, \sigma)$ of the two-time diagram in \cref{fig:8} in the fast-RTD scenario. The TLSs live on a helical quasi-torus, where their position is defined by the angle $2\pi\sigma $ and the length $\theta$. Let us assume for concreteness just two interacting TLSs $S_1$ and $S_2$; the generalization to multiple TLSs is straightforward. Further, since the local time $\sigma$ corresponds to the position on the optical delay line of the RTD-LD circuit, it is reasonable to suppose that the interaction forces between the TLSs do not depend explicitly on the round-trip number $\theta$ but only on the distance $\abs{\sigma_2 - \sigma_1}$ and their ordering in local time. Without loss of generality, assume \mbox{$\sigma_2 > \sigma_1$}.
	
	In general, there can be a "force" forward, $F_{+}$, and backward, $F_{-}$, in local time so that
	\begin{equation}
		\begin{split}
			\tdif{\sigma_1}{\theta} &= F_{-}(\sigma_1 - \sigma_2 + 1) + F_{+}(\sigma_1 - \sigma_2 + 1)\,, \\
			\tdif{\sigma_2}{\theta} &= F_{-}(\sigma_2 - \sigma_1) + F_{+}(\sigma_2 - \sigma_1)\,,
		\end{split}
	\end{equation}
	where adding the period $1$ in the expression for $\dif \sigma_1 / \dif \theta$ accounts for the correct ordering, for $S_2$ can only affect $S_1$ on the next round trip (since causality rules out interaction backward in time $t$). Note that $F_{-}$ and $F_{+}$ are viscous forces because they are proportional to a velocity in the two-time representation rather than an acceleration as is typical of TLSs~\cite{MaggipintoBHF2000,TuraevVZ2012,MunsbergJG2020}.
	
	Assuming an exponential decay with distance because the mechanism of attraction is explained by the slope of the carrier number $n$ (cf.~\cref{sec:results_fastRTD}), the equations of motion are
	\begin{equation}
		\begin{split}
			\tdif{\sigma_1}{\theta} &= - A_{-} \exp(-\gamma_{-} (\sigma_1 - \sigma_2 + 1)T)
			\\&\quad+ A_{+} \exp(-\gamma_{+} (\sigma_1 - \sigma_2 + 1)T)\,, \\
			\tdif{\sigma_2}{\theta} &= - A_{-} \exp(-\gamma_{-} (\sigma_2 - \sigma_1)T)
			\\&\quad+ A_{+} \exp(-\gamma_{+} (\sigma_2 - \sigma_1)T)\,,
		\end{split}
	\end{equation}
	and moreover, we can simplify matters by noting that the attraction forward in time is negligible ($A_+ \ll A_-$) because the tail of the carrier number is small to the left. Thus, the TLSs move only backward in local time
	\begin{equation}
		\begin{split}
			\tdif{\sigma_1}{\theta} &= - A_{-} \exp(-\gamma_{-} (\sigma_1 - \sigma_2 + 1))\,, \\
			\tdif{\sigma_2}{\theta} &= - A_{-} \exp(-\gamma_{-} (\sigma_2 - \sigma_1))\,.
		\end{split}
	\end{equation}
	This tail in $n$ decays exponentially with the rate
	\begin{equation}
		\gamma_{-} = \sub{\gamma}{l} + \sub{\gamma}{m} + \sub{\gamma}{nr}
	\end{equation}
	to the off-state of the laser after a light pulse. The minus sign as index is omitted in the following.
	
	The evolution of the difference $d = \sigma_2 - \sigma_1$ is therefore
	\begin{align}
		\tdif{d}{\theta} &= - A [\exp(-\gamma d T) - \exp(\gamma (d - 1)T)] \notag\\
		&=  - A \exp(-\gamma T / 2)[\exp(-\gamma (d - 1/2)T) \notag\\&\hspace{84pt}- \exp(\gamma (d - 1/2)T)] \notag\\
		&= 2 A \exp(-\gamma T / 2) \sinh(\gamma (d - 1/2)T)\,,
		\label{app:eq:TLS_distance_diffequation}
	\end{align}
	which proves that the equilibrium occurs at $d = 1/2$. To solve differential equation \eqref{app:eq:TLS_distance_diffequation}, we substitute $D = \gamma(d - 1/2)T$ or equivalently $d = D / (\gamma T) + 1/2$ to arrive at
	\begin{equation}
		\begin{split}
			\tdif{D}{\theta}
			&= 2 A \gamma T \exp(-\gamma T/2) \sinh(D) \\
			&= B \sinh(D)
		\end{split}
	\end{equation}
	with $B = 2 A \gamma T \exp(-\gamma T / 2)$.
	Finally, we separate the variables and integrate,
	\begin{align}
		&\frac{1}{B} \INT{D_0}{D}{\tilde{D}} \frac{1}{\sinh(\tilde{D})} 
		= \INT{0}{\theta}{\vartheta} = \theta \notag\\
		&\Leftrightarrow\quad \log\!\bigg(\frac{\tanh(D/2)}{\tanh(D_0/2)}\bigg) = B \theta \notag\\
		&\Leftrightarrow\quad D = 2 \arctanh\!\big(\ee^{B \theta}  \tanh(D_0/2)\big)\,.
	\end{align}
	Resubstituting $d$ for $D$ and recalling that $\gamma = 1/t_\mathrm{n}$, we arrive at the solution
	\begin{align}
		d(\theta) = \frac{1}{2} + 2 \frac{\sub{t}{n}}{T} \arctanh\!\bigg( \ee^{B \theta} \tanh\!\bigg(\frac{T}{2\sub{t}{n}} \bigg[d_0 - \frac{1}{2}\bigg] \bigg) \bigg)\,,
		\label{app:eq:TLS_equation_of_motion}
	\end{align}
	with the fit parameter $B = 2 A \exp\!\big(-\frac{T}{2\sub{t}{n}}\big) \frac{T}{\sub{t}{n}}$.

\end{document}